\documentclass[manuscript]{acmart}
\AtBeginDocument{%
  }

\usepackage{tabularx}
\usepackage[]{mdframed}

\usepackage{adjustbox}

\usepackage{graphicx}
\usepackage{subcaption}

\usepackage{longtable}
\usepackage{lipsum} % just for dummy text- not needed for a longtable
\usepackage[utf8]{inputenc}
\usepackage{pifont}

\usepackage{url}
\usepackage{graphicx}%
\usepackage{multirow}%
\usepackage{newtxmath} % or other font package
\usepackage{amsmath,amssymb,amsfonts}%
\usepackage{amsthm}%
\usepackage{mathrsfs}%
\usepackage[title]{appendix}%
\usepackage{xcolor}%
\usepackage{textcomp}%
\usepackage{manyfoot}%
\usepackage{booktabs}%
\usepackage{algorithm}%
\usepackage{algorithmicx}%
\usepackage{algpseudocode}%
\usepackage{listings}%
\usepackage{graphicx}
\usepackage{caption}
\usepackage{multirow}
\usepackage{float}
\usepackage{soul}
\usepackage{lineno}
\usepackage[flushleft]{threeparttable}
\usepackage{xcolor}
\usepackage{soul}
\usepackage{comment}

\setcopyright{acmlicensed}
\copyrightyear{2018}
\acmYear{2018}
\acmDOI{XXXXXXX.XXXXXXX}
\acmConference[Conference acronym 'XX]{Make sure to enter the correct
  conference title from your rights confirmation email}{June 03--05,
  2018}{Woodstock, NY}
\acmISBN{978-1-4503-XXXX-X/2018/06}

\usepackage{fancyhdr}   %Kopfzeilen und co.
\fancypagestyle{firstpage}{     % define a custom header (Kopfzeile)
  \fancyhf{}
  \chead{\textcolor{gray}{This article has been conditionally accepted for publication in the proceedings of the \\ ACM on Interactive, Mobile, Wearable and Ubiquitous Technologies}}
  \fancyfoot[C]{\small{\textcolor{gray}{~\copyright~ © {Sizhen Bian, et al. | ACM} {2026}. This is a preprint of a work accepted for publication in the ACM Proceedings. It is provided for personal use only and may not be redistributed. Copyright is held by the ACM.}}}

}

\begin{document}

\title{Foundation Models Defining A New Era In Sensor-based Human Activity Recognition: A Survey And Outlook}

% What is a foundational model for HAR?


\author{Sizhen Bian}
\email{sizhen.bian@nwpu.edu.cn}
\affiliation{%
  \institution{Northwestern Polytechnical University}
  %\streetaddress{P.O. Box 1212}
  \country{China}
}

\author{Mengxi Liu}
\email{mengxi.liu@dfki.de}
\affiliation{%
  \institution{DFKI}
  \country{Germany}
}

\author{Lala Shakti Swarup Ray}
\email{}
\affiliation{%
  \institution{RPTU}
  \country{Germany}
}

\author{Bo Zhou}
\email{bo.zhou@dfki.de}
\affiliation{%
  \institution{DFKI}
  \country{Germany}
}

\author{Bin Guo}
\email{guob@nwpu.edu.cn}
\affiliation{%
  \institution{Northwestern Polytechnical University}
  \country{China}
}

\author{Zhiwen Yu}
\email{zhiwenyu@nwpu.edu.cn}
\affiliation{%
  \institution{Northwestern Polytechnical University, Harbin Engineering University}
  \country{China}
}

\author{Thomas Plötz}
\email{thomas.ploetz@gatech.edu}
\affiliation{%
  \institution{Georgia Institute of Technology}
  \country{USA}
}

\author{Paul Lukowicz}
\email{paul.lukowicz@dfki.de}
\affiliation{%
  \institution{DFKI}
  \country{Germany}
}

\author{Siyu Yuan}
\email{syuan@rptu.de}
\affiliation{%
  \institution{RPTU Kaiserslautern-Landau}
  \country{Germany}
}

\author{Vitor Fortes Rey}
\email{fortes@dfki.uni-kl.de}
\affiliation{%
  \institution{DFKI}
  \country{Germany}
}

\renewcommand{\shortauthors}{Sizhen Bian, et al.}

\renewcommand{\shortauthors}{Trovato et al.}

\begin{abstract}
\thispagestyle{firstpage} 
 
Sensor-based Human Activity Recognition (HAR) underpins many ubiquitous and wearable computing applications, yet current models remain limited by scarce labels, sensor heterogeneity, and weak generalization across users, devices, and contexts. Foundation models, which are generally pretrained at scale using self-supervised and multimodal learning, offer a unifying paradigm to address these challenges by learning reusable, adaptable representations for activity understanding. This survey synthesizes emerging foundation models for sensor-based HAR. We first clarify foundational concepts, definitions, and evaluation criteria, then organize existing work using a lifecycle-oriented taxonomy spanning input design, pretraining, adaptation, and utilization. Rather than enumerating individual models, we analyze recurring design patterns and trade-offs across nine technical axes, including modality scope, tokenization, architectures, learning paradigms, adaptation mechanisms, and deployment settings. From this synthesis, we identify three dominant development trajectories: (i) HAR-specific foundation models trained from scratch on large sensor corpora, (ii) adaptation of general time-series or multimodal foundation models to sensor-based HAR, and (iii) integration of large language models for reasoning, annotation, and human–AI interaction. We conclude by highlighting open challenges in data curation, multimodal alignment, personalization, privacy, and responsible deployment, and outline directions toward general-purpose, interpretable, and human-centered foundation models for activity understanding. A complete, continuously updated index of papers and models is available in our companion repository\footnote{\url{https://github.com/zhaxidele/Foundation-Models-Defining-A-New-Era-In-Human-Activity-Recognition}}.

\end{abstract}




\received{20 February 2007}
\received[revised]{12 March 2009}
\received[accepted]{5 June 2009}

\maketitle

\section{Introduction}

% % --- Single-column figure ---
% \begin{figure}[t]
%   \centering
%   \begin{subfigure}[t]{0.49\linewidth}
%     \centering
%     \includegraphics[height=5cm,width=\linewidth]{Figures/HAR_Trend.png}% surveyed works by year
%     \caption{Surveyed HAR-FM papers per year (2021–2025), showing a sharp acceleration with the vast majority of works emerging since 2024.}
%     \label{fig:har-trend}
%   \end{subfigure}\hfill
%   \begin{subfigure}[t]{0.49\linewidth}
%     \centering
%     \includegraphics[height=5cm,width=\linewidth]{Figures/HAR_Cloud.png}% model name cloud
%     \caption{Representative model name cloud, motivating the need for a structured taxonomy and methodological directions.}
%     \label{fig:har-cloud}
%   \end{subfigure}
%   \caption{Growth of the HAR-FM literature and a word cloud of representative model names. 
%\thomas{Figure placement is a bit inconvenient for the reader. Move Fig 2 to later such that the description of Fig 1 is on same page as Fig (ditto for Fig 2)?}
%}
%   \label{fig:trend-and-cloud}
% \end{figure}

\begin{figure}[t]
    \centering
    \includegraphics[width=0.7\linewidth]{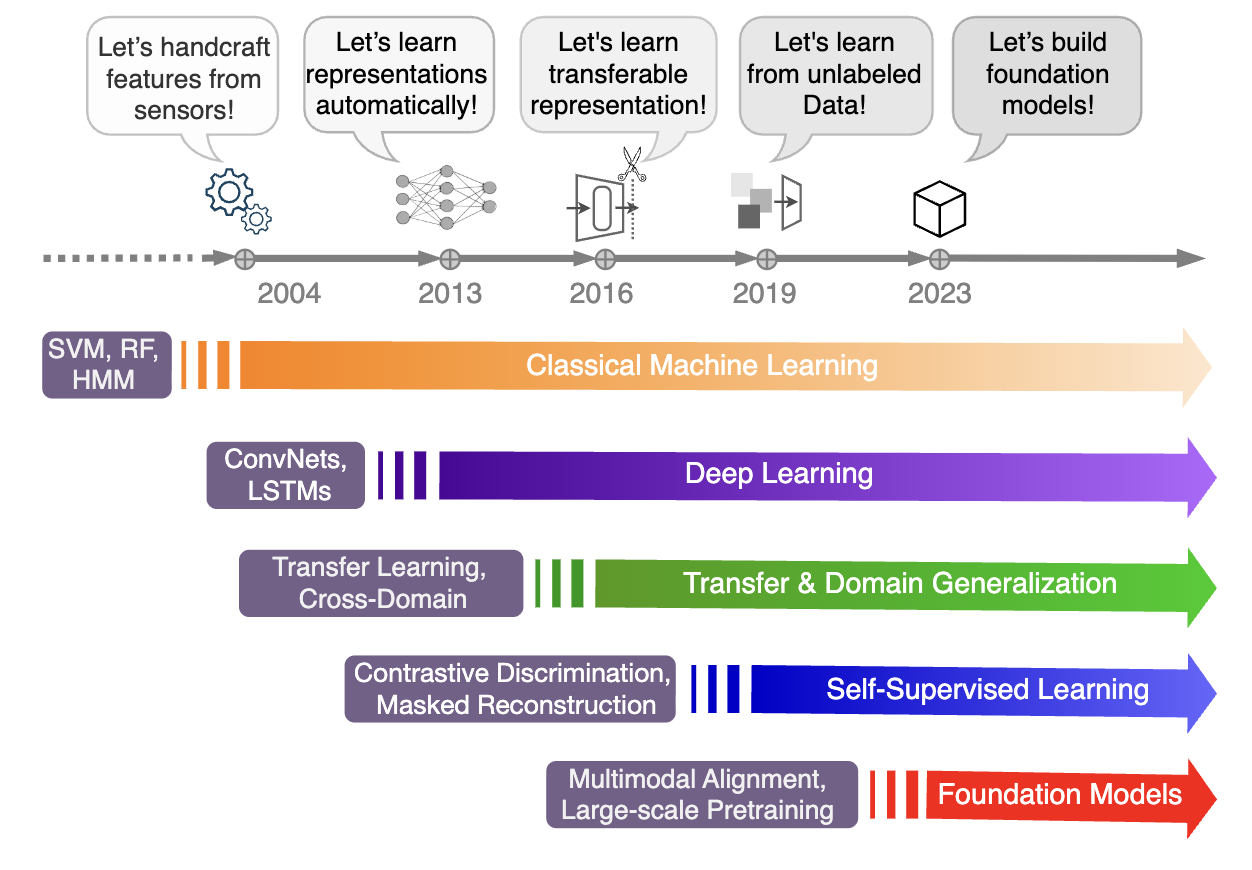}
    \caption{Historical development of sensor-based Human Activity Recognition (HAR) models. 
    From classical machine learning with hand-crafted features and shallow classifiers~\cite{bao2004activity} to the rise of deep learning
    with CNNs and RNNs~\cite{ordonez2016deep}, the field progressed 
    toward a phase focused on transfer and domain generalization (robustness across users, devices, and datasets~\cite{sargano2017human,ramasamy2018recent,wang2018deep}). More recently, self-supervised learning (SSL) 
    approaches have enabled pretraining on unlabeled sensor data using contrastive or masked 
    objectives~\cite{oord2018cpc}. Today, the field 
    is moving toward foundation models, exemplified by large-scale sensor--language 
    alignment~\cite{sensorlm2025}, emphasizing scalability, generalization, and interpretability.}
    \label{fig:har_history_timeline}
\end{figure}

Sensor-based Human Activity Recognition (HAR) has long been a cornerstone of ubiquitous and wearable computing, enabling applications in health monitoring \cite{alam2024neurohar,kwon2021approaching}, human–computer interaction \cite{chen2021deep, bian2024body}, and smart environments \cite{akila2023human, kalimuthu2021human}.  
Despite the rapid evolution of deep learning, most HAR models remain limited by their reliance on task-specific architectures, small annotated datasets, and poor generalization across users, devices, or contexts \cite{arshad2022human, antar2019challenges, islam2022human,karim2024human}.  
Recently, the emergence of foundation models has suggested a viable route to mitigate HAR’s core constraints, fundamentally transforming how activity understanding can be approached from wearable, ambient, and physiological sensor data \cite{leng2023benefit, qian2024advancing, zhang2025sensorlm}. As illustrated in Figure \ref{fig:har_history_timeline}, the development of sensor-based HAR has followed a clear trajectory, from classical machine learning based on handcrafted features, through deep, transferable, and self-supervised learning, to today’s emerging foundation models. This evolution reflects a continuous pursuit of scalability, generalization, and interpretability in understanding human activity from sensor data \cite{dosovitskiy2020image, ji2024hargpt}. 
%Unlike traditional machine-learning pipelines that depend heavily on narrow supervised training, foundation models leverage large-scale pretraining on diverse, often unlabeled data to learn transferable representations that can be efficiently adapted to specific downstream tasks \cite{dosovitskiy2020image, ji2024hargpt}. 

The rising interest in sensor-based HAR foundation models stems from several converging developments:  
(1) the proliferation of constantly emerging wearable and ambient datasets capturing multimodal physiological, kinematic, and contextual signals \cite{yuan2024self, saha2025pulse, hoddes2025scaling};  
(2) advances in self-supervised learning techniques such as masked autoencoding and contrastive alignment \cite{haresamudram2022assessing, jain2022collossl, cheng2024maskcae};  
(3) the adaptation of transformer architectures to sequential and multimodal sensor streams \cite{xiao2022two, islam2020hamlet,shavit2021boosting}; and  
(4) the emergence of cross-modal and sensor–language models that bridge raw sensor signals and semantic concepts \cite{zhang2024unimts, liu2023large, leng2023benefit, driess2023palm, liu2023biosignal,moon2023imu2clip}.  
These trends collectively extend the scaling behaviors and representational richness once confined to vision and language domains into the realm of sensor-based HAR.

Motivated by these developments, this survey provides the first comprehensive synthesis of foundation models for sensor-based HAR. 
While prior surveys, such as \cite{haresamudram2025past, bian2022state}, have reviewed the general evolution and trends of sensor-based HAR, they have not examined foundation models as a significantly distinct paradigm nor analyzed their emerging design principles and domain-specific requirements. 
The earliest work closely aligned with the concept of foundation models in sensor-based HAR is LIMU-BERT \cite{xu2021limu}, introduced in 2021. This pioneering model leveraged self-supervised learning to extract generalized, task-agnostic representations from unlabeled IMU data, inspired by the pretraining principles of the BERT language model to effectively capture temporal relations and statistical distributions within inertial signals. In 2022, the pretrain-then-finetune paradigm grounded in self-supervised learning gained rapid momentum as researchers sought to mitigate the data scarcity challenge in sensor-based HAR \cite{haresamudram2022assessing}. These early efforts paved the way for the current wave of foundation models by emphasizing label-efficient learning, multimodal alignment, and transferable representations. As a result, the community has observed a steep upward trend in foundation models for sensor-based HAR, with the majority of studies emerging since 2024. 

This work analyzes 132 representative papers and organizes the field using a unified, lifecycle-oriented taxonomy spanning input design, pretraining, adaptation, and utilization, structured across nine technical and operational axes. Building on this taxonomy, this work identifies three dominant development directions: 
(1) developing HAR-specific foundation models from scratch, 
(2) adapting general time-series and multimodal foundation models to sensor data, and 
(3) leveraging large language models for reasoning, annotation, and human--AI interaction. 
Open challenges are further discussed, including large-scale data curation, multimodal alignment, on-device personalization, and responsible deployment, and future research directions are outlined toward robust, generalizable, and human-centered foundation models for activity understanding.

It should be noted that this work focuses specifically on sensor-based HAR and deliberately excludes vision- or video-based HAR from the scope of this survey. Here, sensor-based HAR refers to recognition approaches based on non-visual sensing modalities, including inertial sensors (e.g., accelerometers and gyroscopes), physiological sensors (e.g., ECG, PPG), RF signals, acoustic sensing, and other ambient or wearable sensing technologies. Vision-based HAR, including approaches relying on RGB/depth cameras, wearable cameras, and vision–language models (VLMs), represents a distinct and well-established research direction with its own datasets, benchmarks, and modeling paradigms. While such methods are highly relevant to the broader field of ubiquitous computing, they differ fundamentally from sensor-based HAR. In particular, challenges such as data scarcity and annotation complexity are generally less severe in vision-based approaches due to the availability of large-scale datasets and mature benchmarks, whereas sensor-based HAR often operates with limited, heterogeneous, and less standardized data. As a result, the two paradigms involve different data characteristics, modeling assumptions, and system design considerations. Including both within a single survey would blur these differences and hinder a coherent and in-depth analysis. Therefore, this survey focuses on sensor-based HAR to maintain clarity and depth, while treating vision-based methods only as auxiliary components when relevant (e.g., for cross-modal alignment or semantic supervision), rather than as primary sensing modalities for HAR.

\section{Background and Motivation}
\label{sec:background}

\subsection{Sensor-based Human Activity Recognition: Foundations and Challenges}
\label{subsec:challenges}

Sensor-based HAR seeks to infer human actions, behaviors, and internal states from sensor data~\cite{chen2021deep, wang2019deep, vaizman2018context, bian2022exploring, bian2022state, zhou2019widar3}. 
It spans a broad range of sensing modalities, including inertial measurement units (IMUs)~\cite{zhou2025imucoco, bian2025hybrid}, radio-frequency (RF) sensing (e.g., WiFi, mmWave radar)~\cite{ma2025adapttrack, guo2025mmpencil}, physiological signals such as photoplethysmography (PPG) and electrodermal activity (EDA)~\cite{fedorin2025virtual, meier2024robust, alchieri2025improving, mercado2024eda}, and emerging modalities such as capacitive, magnetic, acoustic, and hybrid sensors~\cite{liu2025facial, giordano2025pulse, bian2020wearable, bian2024body, gaya2024deep, bonazzi2023tinytracker, bonazzi2024retina}. 
Each modality captures complementary aspects of human behavior (e.g., body motion, spatial displacement, physiological and affective states), making Sensor-based HAR inherently multimodal and context-dependent.
Despite sustained progress, sensor-based HAR remains fundamentally distinct from perception domains such as vision or speech. 
As summarized in Fig.~\ref{fig:har_challenges}, its challenges arise from multiple abstraction levels:

\begin{figure}[!htb]
    \centering
    \includegraphics[width=0.85\linewidth]{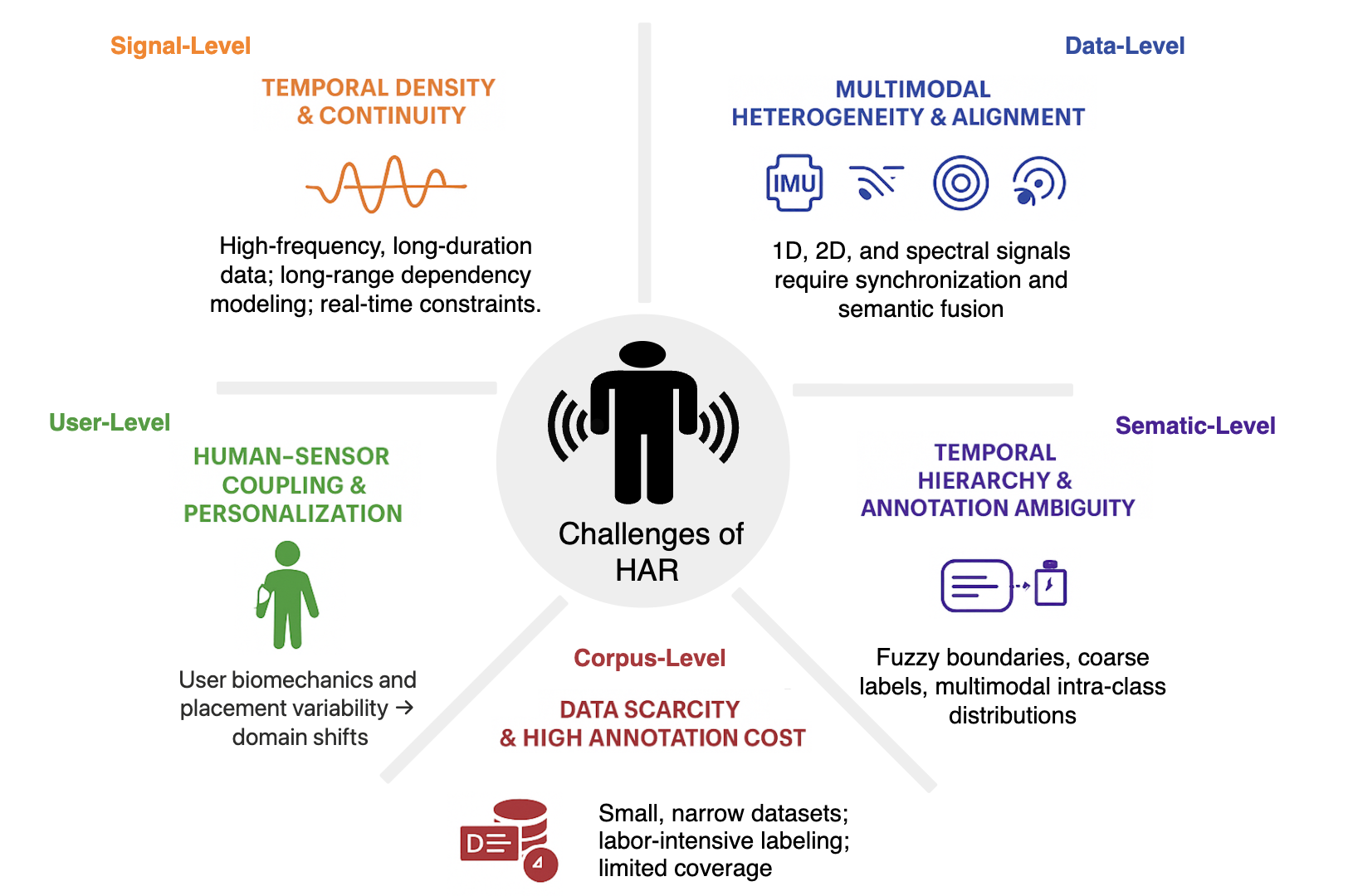}
    \caption{
    Foundations and challenges of sensor-based HAR across multiple abstraction levels:
    signal-, data-, user-, semantic-, and corpus-level factors jointly define the complexity of learning robust and generalizable activity representations.
    }
    \label{fig:har_challenges}
\end{figure}

\textbf{Signal-level challenges} stem from the continuous, high-frequency, and body-coupled nature of sensor streams. Wearable and ambient devices often sample at tens to hundreds of hertz, producing long sequences that encode fine-grained motion and physiological dynamics~\cite{yang2015deep, niazi2017statistical}. Capturing long-range temporal dependencies, handling irregular sampling or dropouts, and meeting real-time constraints (latency, energy, reliability) remain persistent difficulties, particularly for deployment-oriented systems.

\textbf{Data-level challenges} arise from modality heterogeneity in both structure and semantics. While inertial and physiological sensors generate one-dimensional time series, RF, radar, and acoustic sensing produce spatial or spectral representations. Effective HAR requires both temporal synchronization of asynchronous streams and semantic alignment across modalities, mapping disparate signals to shared latent concepts such as activity intensity, intent, or context. Conventional fusion strategies (e.g., late fusion or handcrafted alignment) often fail to capture these deeper cross-modal relationships, limiting robustness and interpretability.

\textbf{User-level challenges} reflect the strong coupling between sensing devices and individuals. Sensor signals vary substantially across users due to biomechanics, placement, orientation, clothing, and device characteristics~\cite{ramanujam2021human, bian2021systematic}. Even within a single user, temporal drift arises from fatigue, sensor repositioning, or contextual changes. These effects induce domain shift and concept drift across users, devices, and time~\cite{stisen2015smart, guan2017ensembles, kang2024device}. While personalization can mitigate such variability, it typically requires labeled data and frequent retraining, creating a tension between scalability and individual adaptation.

\textbf{Semantic-level challenges} originate from the hierarchical and ambiguous nature of human activities. Activities unfold over variable temporal scales, exhibit fuzzy boundaries, and form nested structures (e.g., gestures within actions, actions within routines)~\cite{cheng2013learning, kuehne2014language}. Labels are often coarse, inconsistent, or weakly defined, and intra-class variability is high due to style and context~\cite{kwon2019handling, demrozi2023comprehensive, barshan2016investigating}. As a result, window-based classification struggles to capture transitions, long-term context, and higher-level intent.

\textbf{Corpus-level challenges} remain a systemic bottleneck. Unlike vision or language, sensor-based HAR lacks web-scale, standardized datasets. Most corpora are small, curated, and context-specific, with costly annotation pipelines involving manual labeling, video synchronization, or self-reports~\cite{roggen2010collecting, demrozi2021towards}. Dataset heterogeneity in label definitions, sensor setups, and protocols further hinders reproducibility and cross-dataset generalization~\cite{benchekroun2023cross, zhang2024reproducible}. Consequently, rare activities, long-tail behaviors, and real-world variability are poorly represented.

\paragraph{Takeaway.}
Sensor-based HAR poses a uniquely demanding setting for representation learning, where robustness, adaptability, and cross-context generalization are as critical as raw recognition accuracy. These challenges motivate modeling paradigms that can exploit large-scale unlabeled data, integrate heterogeneous modalities, and support flexible adaptation, setting the stage for foundation models tailored to the requirements of human activity understanding. The next subsection formalizes what constitutes a foundation model in the sensor-based HAR context and explains why this paradigm is a promising response to the challenges outlined above.

\subsection{Foundation Models and Their Relevance to Sensor-based Human Activity Recognition}
\label{subsec:definition}

\paragraph{Definition of FM in sensor-based HAR}

\begin{figure}[!htb]
  \centering
  \includegraphics[height=5.7cm,width=0.99\linewidth]{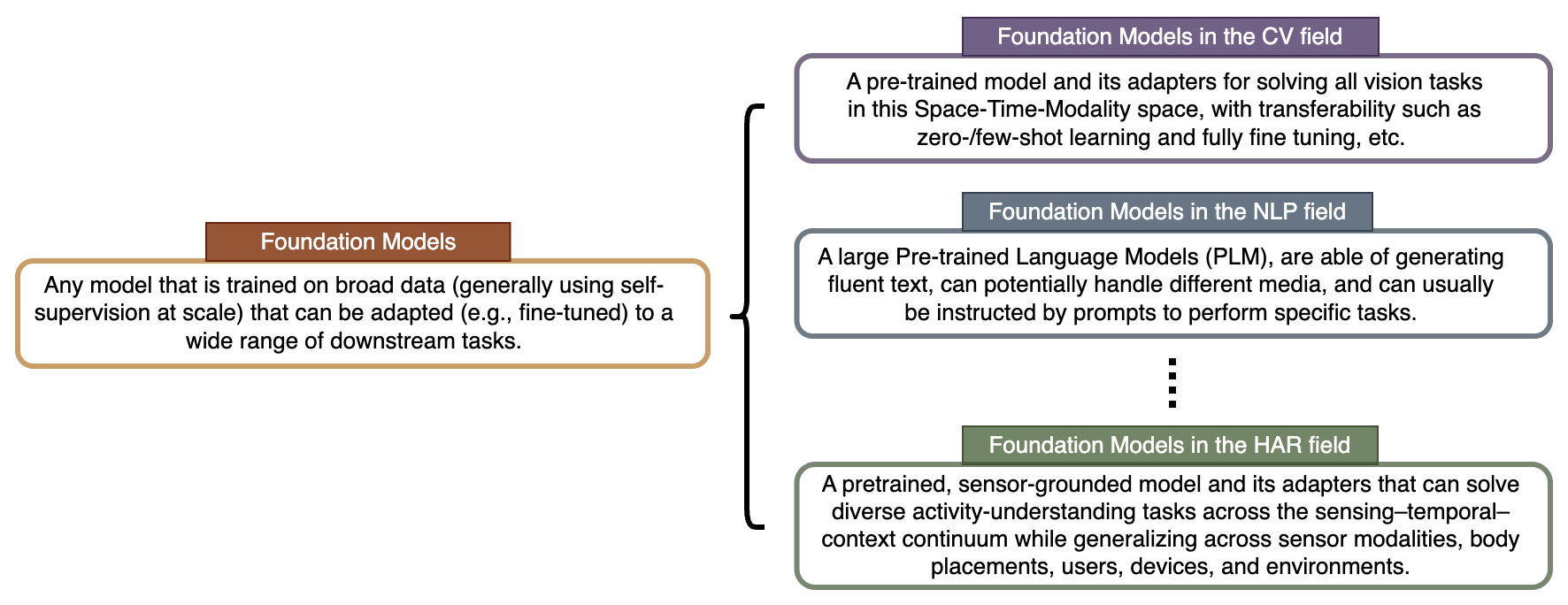}
  \caption{Definition of Foundation Models \cite{wiggins2022opportunities} and its adaptation across Computer Vision \cite{yuan2021florence}, Natural Language Processing \cite{paass2023foundation}, and Sensor-based Human Activity Recognition (this work).}
  \label{fig:har-definition}
\end{figure}

The term Foundation Model was introduced by the Stanford Center for Research on Foundation Models (CRFM) to describe models trained on broad data (typically using self-supervision at scale) that can be adapted to a wide range of downstream tasks \cite{wiggins2022opportunities}. As the paradigm matured, domain-specific interpretations emerged. As Figure \ref{fig:har-definition} illustrates, in computer vision, foundation models are viewed as pretrained backbones with adapters spanning a space–time–modality continuum \cite{yuan2021florence}; In natural language processing, the term is largely reserved for large pretrained language models capable of instruction following and generative reasoning \cite{paass2023foundation}.
Sensor-based HAR differs fundamentally from both domains: activities are realized through diverse sensing modalities, unfold over time, and are shaped by embodiment, context, and environment. Consequently, a foundation model for sensor-based HAR must generalize not only across tasks, but also across sensors, users, and deployment conditions. We therefore adapt the notion of foundation models to sensor-based HAR as follows:

\vspace{4mm}
\begin{mdframed}
A \textbf{Foundation Model in Sensor-based Human Activity Recognition} is a pretrained, sensor-grounded model and its adapters that can solve diverse activity-understanding tasks across the sensing–temporal–context continuum while generalizing across sensor modalities, body placements, users, devices, and environments.
\end{mdframed}
\vspace{4mm}

In practice, sensor-based HAR foundation models are typically trained with self-supervised objectives on large-scale, heterogeneous sensor corpora (wearable, ambient, RF, physiological), optionally incorporating language or vision for cross-modal grounding. They expose stable interfaces (via embeddings, prompts, or lightweight adapters) that enable zero-/few-shot transfer, multimodal fusion, and efficient personalization without task-specific retraining. These models operate across (i) diverse sensing modalities, (ii) temporal hierarchies ranging from micro-events to long-term routines, and (iii) contextual and embodiment variations such as placement, device form factor, and user diversity.

\begin{figure}[t]
    \centering
    \includegraphics[width=0.60\linewidth]{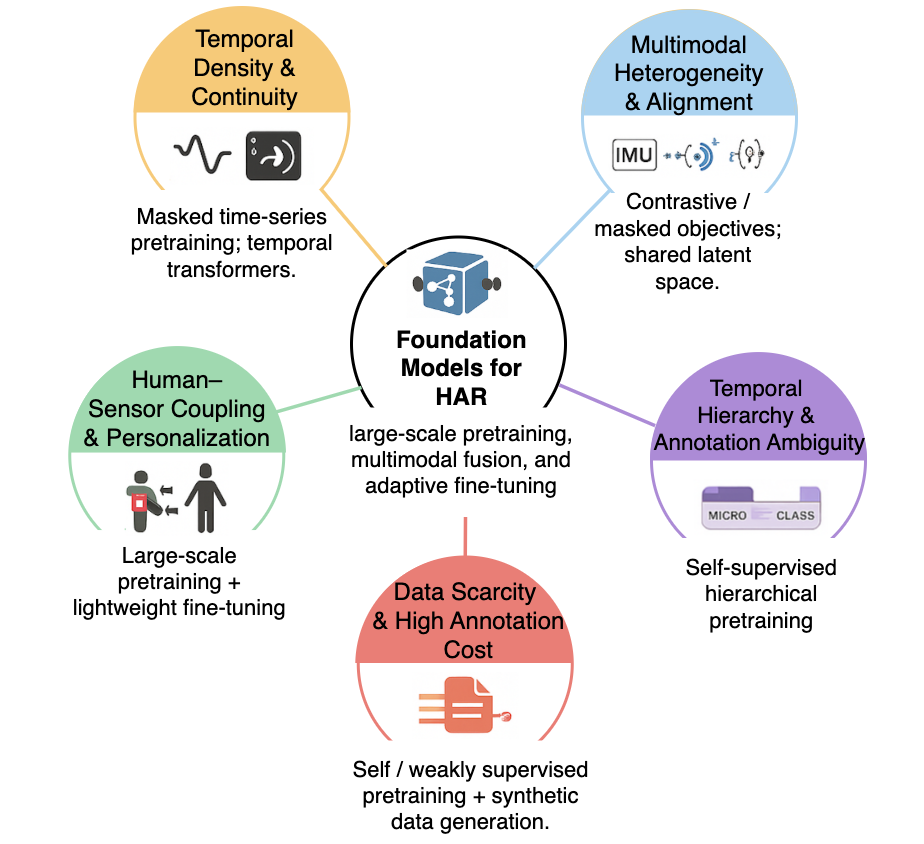}
    \caption{How foundation models address HAR challenges across signal-, data-, user-, semantic-, and corpus-level dimensions.}
    \label{fig:fm_har_solutions}
\end{figure}

Having established this definition, we next explain why sensor-based HAR is a particularly suitable domain for the foundation model paradigm. 
%Foundation models have reshaped AI by enabling scalable representation learning through large-scale pretraining \cite{bommasani2021opportunities, schneider2024foundation}. 
In vision and language, architectures such as Vision Transformers and large language models demonstrate that self-supervised objectives can yield transferable, task-agnostic representations \cite{dosovitskiy2020image, brown2020language, touvron2023llama}. These successes motivate extending the paradigm to multimodal time-series domains, including sensor-based HAR \cite{zhang2025sensorlm, narayanswamy2025scaling}. Building on the challenges identified in Subsection~\ref{subsec:challenges}, sensor-based HAR closely matches the conditions under which foundation models have proven effective: abundant unlabeled data, heterogeneous inputs and the need for generalization. Figure~\ref{fig:fm_har_solutions} summarizes how foundation models address HAR challenges across multiple abstraction levels.

\textbf{Signal-level (temporal density and continuity).}
Sensor-based HAR signals are continuous and high-frequency, making long-range temporal modeling difficult for window-based or shallow sequential models \cite{liu2024timer, miller2024survey}, e.g., in capturing transitions between activities and temporal correlations across multiple time scales. Foundation models mitigate this by leveraging masked time-series pretraining and temporal transformers, which learn hierarchical temporal abstractions over long horizons via reconstruction and prediction objectives \cite{liang2024foundation, ahmadi2025unsupervised, ye2024survey}, offering a more principled approach to handling temporal density and continuity.

\textbf{Data-level (multimodal heterogeneity and alignment).}
Sensor-based HAR spans modalities with disparate structures and semantics. Foundation models learn shared latent spaces through contrastive or masked multimodal objectives, enabling semantic alignment across asynchronous sensor streams \cite{chen2024evolution, zheng2024heterogeneous}, thereby moving sensor-based HAR toward a holistic understanding of human behavior and bridging the physical, physiological, and contextual layers of activity sensing. Analogous to image–text alignment in CLIP \cite{radford2021learning}, sensor foundation models such as SensorLM \cite{sensorlm2025} support cross-modal generalization and zero-shot multimodal reasoning.

\textbf{User-level (human–sensor coupling and personalization).}
Large-scale pretraining exposes models to diverse users, placements, and devices, yielding population-level representations robust to individual variability \cite{chen2024large}. Lightweight adaptation mechanisms (few-shot fine-tuning, on-device learning, or prompt conditioning) enable efficient personalization without full retraining \cite{geng2023personalized, lee2025tap}, bridging generalization and individual specificity \cite{kang2025bridging}. Some recent approaches also explore continual and federated adaptation, allowing user devices to locally refine shared foundation models while maintaining global coherence across populations.

\textbf{Semantic-level (temporal hierarchy and annotation ambiguity).}
Activities exhibit nested temporal structure and weak, inconsistent labels. Self-supervised objectives such as masked reconstruction and temporal discrimination allow foundation models to discover multi-scale semantic structure directly from unlabeled sequences \cite{zhang2024self, jaiswal2020survey, ye2023hitea}, effectively discovering multi-scale temporal semantics from instantaneous gestures to extended activity routines. This improves label efficiency and robustness to annotation noise while supporting hierarchical activity reasoning.

\textbf{Corpus-level (data scarcity and annotation cost).}
Foundation models decouple representation learning from manual labeling by exploiting large volumes of unlabeled sensor data \cite{ericsson2022self, kimura2024vibrofm, zhang2025sensorlm}. Through large-scale pretraining objectives, such as masked reconstruction, temporal contrastive learning, or multimodal co-alignment, foundation models can extract behavioral structure from raw data without explicit human supervision. Emerging generative and language-grounded approaches further augment scarce datasets through synthetic data and sensor–text alignment \cite{leng2023benefit, sharma2025sensorgpt, zhou2024advancing, sharma2025synthetic}.

\paragraph{Takeaway.}
Foundation models offer a coherent response to the intertwined challenges of sensor-based HAR by unifying large-scale representation learning, multimodal alignment, and efficient adaptation. Rather than optimizing isolated tasks or datasets, they provide reusable backbones that support robust, personalized, and semantically grounded activity understanding across diverse real-world settings.

\subsection{Criteria for Foundation Models in Human Activity Recognition}
\label{subsec:criteria}

Translating the foundational model to sensor-based HAR necessitates consideration of the domain’s unique characteristics. Building on the definition introduced in Subsection~\ref{subsec:definition}, we articulate a set of diagnostic criteria that characterize whether a model can reasonably be regarded as a foundation model for sensor-based HAR. These criteria are not a rigid checklist; rather, they capture recurring design principles observed across emerging HAR foundation models and reflect the degree to which a model supports reuse, generalization, and scalable adaptation.

\begin{figure*}[!hbt]
  \centering
  \includegraphics[width=0.85\textwidth]{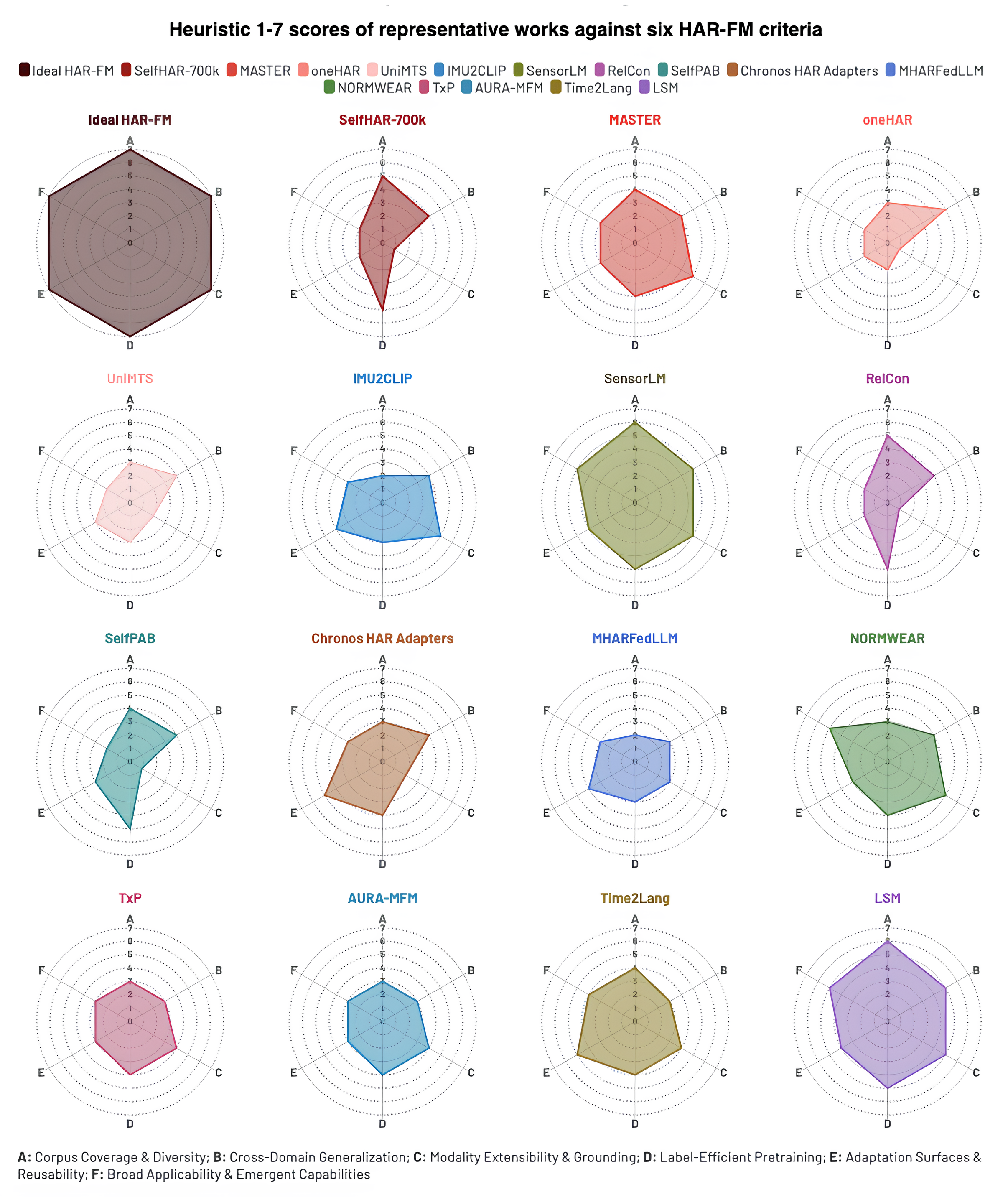}
  \caption[Heuristic scores of representative works against HAR-FM criteria]{
  \textbf{Heuristic 1--7 scores of representative works against six HAR–FM criteria.}
%  Each radar chart profiles a model on:
%  \textbf{A)} Corpus Coverage \& Diversity,
%  \textbf{B)} Cross-Domain Generalization,
%  \textbf{C)} Modality Extensibility \& Grounding,
%  \textbf{D)} Label-Efficient Pretraining,
%  \textbf{E)} Adaptation Surfaces \& Reusability, and
%  \textbf{F)} Broad Applicability \& Emergent Capabilities.
  The “Ideal HAR-FM” panel depicts a target profile.
  Scores (1\,{=}\,limited evidence to 7\,{=}\,strong evidence) are judgment-based syntheses from reported results (compared both to the other models in this survey and to an aspirational “ideal” FM-for-HAR reference point) and are intended for qualitative comparison rather than a leaderboard. Representative works: SelfHAR-700k~\cite{yuan2024self}, MASTER \cite{zhu2024master}, OneHAR~\cite{Wei_2025}, UniMTS~\cite{zhang2024unimts}, IMU2CLIP~\cite{moon2023imu2clip}, SensorLM~\cite{zhang2025sensorlm}, RelCon~\cite{xu2024relcon}, Chronos HAR Adapters~\cite{xiong2024novel}, MHARFedLLM~\cite{bandyopadhyay2025mharfedllm}, NORMWEAR~\cite{luo2024normwear}, TxP \cite{ray2025txp}, AURA-MFM~\cite{matsuishi2025multimodal}, Time2Lang~\cite{pillai2025time2lang}, and LSM~\cite{narayanswamy2024scaling}.
  }
  \label{fig:har_criteria_radar}
\end{figure*}

\begin{enumerate}

\item \textbf{Corpus Coverage \& Diversity.}
Foundation models in the sensor-based HAR domain should encompass corpora that span users, devices, placements, settings, and temporal regimes. In practice, this breadth often correlates with scale and diversity; robust generalization typically requires both large corpora and heterogeneous sources (multi-dataset, multi-device, multi-context)\cite{yuan2024self, zhu2024master}. 

\item \textbf{Cross-Domain Generalization.}  
A defining property of foundation models is their ability to generalize across users, devices, and datasets, including zero-shot and few-shot transfer.  In practice, this means maintaining performance under non-IID shifts, temporal drift, and open-set conditions. ~\cite{Wei_2025, zhang2024unimts}

\item \textbf{Modality Extensibility \& Grounding.}  
Foundation models should scale across sensing channels by either (a) natively encoding multiple modalities within a unified backbone, or (b) exposing stable attachment points (projection heads, shared embedding spaces, or cross-attention bridges) so new modalities can be added with minimal architectural churn. In practice, this means tolerating variable sampling rates, remaining robust to missing channels, and aligning sensor representations to external semantics~\cite{moon2023imu2clip, zhang2025sensorlm}.

\item \textbf{Label-Efficient Pretraining.}
While not strictly mandatory, HAR FMs typically rely on self-supervised or hybrid objectives (masked prediction, contrastive alignment, generative reconstruction) to exploit unlabeled streams and reduce dependence on costly annotations\cite{xu2024relcon, logacjov2024selfpab} .

\item \textbf{Adaptation Surfaces \& Reusability.}
A pretrained backbone, either a sensor encoder or a general-purpose LLM, should expose lightweight adaptation interfaces (adapters/LoRA, projection heads, prompt/prefix tuning) that enable rapid retargeting to new tasks with small trainable-parameter budgets~\cite{xiong2024novel, bandyopadhyay2025mharfedllm}.

\item \textbf{Broad Applicability \& Emergent Capabilities.}
HAR foundation models should expose general-purpose representations and flexible interfaces so that a single pretrained core can be composed into diverse analytic and interactive workflows spanning health monitoring, behavioral analytics, multimodal retrieval, and language-grounded explanation. Beyond tasks seen during training, strong representations should also exhibit emergent behaviors, delivering reasonable performance on related, previously unseen tasks (e.g., inferring step counts from walking embeddings, user/placement recognition, or demographic proxies) via prompting, probing, or light adaptation~\cite{luo2024normwear}. 
    
\end{enumerate}

Although foundation models in the sensor-based HAR domain are still in their formative phase, a model should exhibit core hallmarks of the foundation model paradigm, not necessarily all at once, but in substance. 
Figure~\ref{fig:har_criteria_radar} translates these criteria into radar profiles of representative FM-HAR systems, providing a diagnostic lens rather than a ranking. Each shaded polygon (scored heuristically from 1–7 across data scale/diversity, non-IID protocols, modality handling, pretraining objectives, adaptation interfaces, and task breadth) captures a distinct design philosophy: some emphasize large, heterogeneous corpora and label-efficient pretraining; others focus on reusable adaptation surfaces or multimodal grounding; a few pursue broad task coverage across health, daily living, and retrieval. The “Ideal HAR-FM” outline marks more than an aspirational target; it defines a strategic roadmap for the field.
The gaps and asymmetries across profiles expose where the next breakthroughs must occur: scaling datasets beyond controlled settings, integrating heterogeneous sensors within unified embedding spaces, and developing lightweight yet expressive adaptation mechanisms that preserve privacy and on-device efficiency.
Collectively, the radar plots function as guidelines for progress, highlighting that the path forward lies not in a single dimension of improvement, but in converging broad pretraining, multimodal grounding, and adaptive personalization under standardized non-IID evaluation protocols.

\paragraph{LLMs as foundation models in sensor-based HAR}
We regard a model as a foundation model, especially targeting sensor-based HAR, when it can be reused across tasks, cohorts, and datasets with minimal task-specific supervision and exposes a stable, general-purpose interface for adaptation. %\thomas{Finally some sort of a definition. Why so late?} 
Under this lens, LLMs qualify as HAR FMs in three practical forms: (i) \emph{sensor-anchored FMs with an LLM decoder} (encoder--decoder stacks) in which learned sensor tokenizers/encoders condition a language decoder via cross-attention~\cite{chen2024sensor2text,zhang2025sensorlm}; 
%(e.g., \textbf{Sensor2Text}~\cite{chen2024sensor2text}, \textbf{SensorLM}~\cite{zhang2025sensorlm}); 
(ii) \emph{sensor--language FMs} (LM stacks) in which a frozen or partially frozen LLM directly ingests fixed-length sensor tokens or projections for captioning, retrieval, and reasoning~\cite{pillai2025time2lang,li2025sensorllm}
%(e.g., \textbf{Time2Lang}~\cite{pillai2025time2lang}, \textbf{SensorLLM}~\cite{li2025sensorllm}); 
and (iii) \emph{prompt-only, in-context LLMs} that operate with a deterministic sensor$\rightarrow$prompt pipeline (e.g., windows, summaries, or event streams formatted as text) and demonstrate consistent zero-/few-shot transfer across multiple HAR tasks without weight updates~\cite{thapa2025stressllm,ji2024hargpt}.
%(e.g., \textbf{StressLLM}~\cite{thapa2025stressllm}, \textbf{HARGPT} \cite{ji2024hargpt}). 
By contrast, the ad hoc use of a generic LLM with hand-crafted prompts for a single dataset, without a reproducible sensor-to-prompt interface or evidence of cross-task reuse, remains an auxiliary reasoning component rather than an HAR FM.

\subsection{Survey Methodology and Scope}
\label{subsec:survey}

To ensure comprehensive coverage of recent developments in foundation models in HAR domain, we conducted a multi-source literature search spanning peer-reviewed journals, flagship conferences, and preprint repositories. The search drew upon Google Scholar and Scopus for broad bibliographic retrieval, supplemented by domain-specific sources such as arXiv and the SciSpace corpus, as well as leading venues including NeurIPS, AAAI, CVPR, UbiComp/ISWC, MobiSys, MobiCom, and journals such as IEEE TPAMI, IEEE TNNLS, IEEE JBHI, and ACM IMWUT.
The search window covered publications up to late~2025. In total, we collected 132 papers closely aligned with our objectives.

\textbf{Search strategy.} We combined keywords such as \emph{human activity recognition}, \emph{foundation model}, \emph{sensor}, \emph{time-series}, \emph{self-supervised}, \emph{multimodal}, and \emph{language model}. We retained the most recent version when multiple preprint revisions existed and de-duplicated overlapping arXiv/conference versions. In addition, we systematically tracked both the references cited within each paper and the subsequent works citing it, ensuring comprehensive coverage of the evolving research landscape.

\textbf{Screening and eligibility.} A two-stage screening (title/abstract, then full text) applied the following inclusion criteria: (i) explicit focus on foundation-model style approaches to sensor-based HAR; (ii) a substantive architectural or methodological contribution (e.g., pretraining, adaptation, cross-/multimodal alignment); and (iii) empirical evaluation on established HAR datasets or diverse sensing modalities. We excluded purely vision-only works with no sensor grounding and classical pipelines without FM components.

%\textbf{Data extraction and coding.} Each included paper was coded against a unified, four-phase schema that we use throughout the survey: \emph{Prepare} (modality scope; data landscape; tokenization \& representation), \emph{Pretrain} (base architecture and pretraining paradigm), \emph{Adapt} (parameter-efficient\ full/partial fine-tuning, etc.), and \emph{Deploy} (downstream capabilities; practical deployment; and application domains). Note that not all FM–HAR works span all phases: some use a preexisting backbone without new pretraining (Adapt), and many report representation quality without a deployment study (Deploy); accordingly, we coded only the phases for which methods and experiments were explicitly documented.

\section{Taxonomy of Foundation Models in Sensor-based HAR}
\label{sec:taxonomy}

%-------------------------------------------
% Figure: Four-stage pipeline taxonomy (Design → Pretraining → Adaptation → Application)
%-------------------------------------------
\begin{figure*}[b]
  \centering
  % Replace the filename with your final asset path/name
  \includegraphics[width=\textwidth]{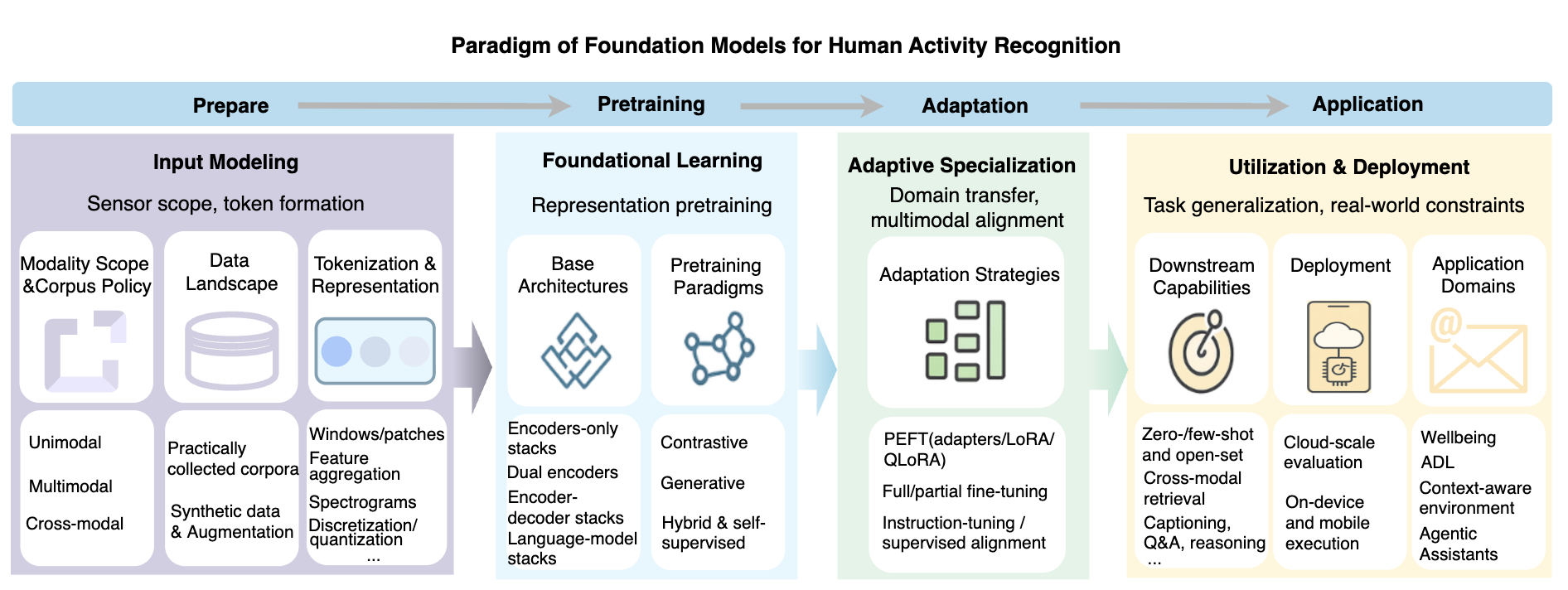}
  \caption{ \textbf{Conceptual taxonomy and lifecycle of foundation models for sensor-based Human Activity Recognition (HAR).}
  This framework synthesizes the diverse methodological patterns into four–phase workflow: 
  %\emph{Design} (Input Modeling) fixes what the model will see by choosing the \emph{modality scope \& corpus design}, by preparing the \emph{data landscape}, and by shaping signals into learnable units through \emph{tokenization \& representation}. \emph{Pretraining} (Foundational Learning) couples \emph{base architectures} with \emph{pretraining paradigms} to acquire broad, transferable priors over these tokens. \emph{Adaptation} (Adaptive Specialization) then tailors those priors to new users, devices, and tasks, and performs multimodal fusion/alignment, typically with parameter–efficient methods or selective fine–tuning, including federated variants. Finally, \emph{Application} (Utilization \& Deployment) treats the model as a capability evaluated in real tasks: downstream probes (zero-/few-shot and open-set recognition, cross-modal retrieval, captioning/Q\&A/reasoning, and reconstruction/forecasting) are assessed (sometime alongside practical deployment considerations) across target domains (ADL/general-purpose, healthcare including sleep, smart home/context, interactive \& agentic assistants). 
  By aligning these dimensions along the model development lifecycle, the taxonomy clarifies how individual architectural or methodological choices contribute to the progression from general pretraining to contextual grounding and real-world application, thereby offering a unifying structure for subsequent analysis in Section~\ref{sec:taxonomy}. 
  }
  \label{fig:fm_har_taxonomy}
\end{figure*}

We recast the development of HAR foundation models as a four–phase pipeline. 
(\emph{i}) \textbf{Prepare (Input Modeling)}: decide the modality scope and corpus design, prepare the data landscape, and shape continuous streams into learnable units. 
(\emph{ii}) \textbf{Pretraining (Foundational Learning)}: couple base architectures with contrastive, generative, or hybrid self-supervised objectives to acquire broad, transferable sensor priors over these tokens. 
(\emph{iii}) \textbf{Adaptation (Adaptive Specialization)}: tailor those priors to new users, devices, and tasks through parameter-efficient, selective/full layer-wise, or instruction-based fine-tuning, and perform modality fusion/alignment. 
(\emph{iv}) \textbf{Application (Utilization \& Deployment)}: exercise the pretrained backbones for zero-/few-shot and open-set recognition, retrieval and language grounding (captioning, Q\&A, reasoning), and reconstruction/forecasting under real-world constraints (cloud / on-device). 
This pipeline makes explicit how early choices about sensor scope and token design propagate through learning and adaptation to determine downstream capability and deployability. Figure~\ref{fig:fm_har_taxonomy} summarizes nine complementary axes that cut across these stages.
Together, these dimensions provide a multi-faceted view of how foundation models evolve from data-centric pretraining to system-level generalization. They also bridge conceptual and practical perspectives: the first three dimensions focus on \emph{what the model sees and at what resolution}, the second two dimensions focus on \emph{how models learn}, the middle one on \emph{what they represent}, and the final three on \emph{how they are used}. This structure forms the basis of the following subsections, each analyzing one design axis in detail, grounded in representative examples from the literature summary.

To be noticed, as mentioned in \ref{subsec:survey}, we use the lifecycle as an organizing lens rather than a checklist: individual works need not span every phase to count as FM-for-HAR contributions. In practice, a study may legitimately focus on a specific stage (e.g., large-scale pretraining, few-shot adaptation, on-device optimization, or post-deployment monitoring) without covering the entire lifecycle. 
%(as Figure \ref{fig:lifecycle_matrix} illustrates with a few representative works).

%\begin{figure}[t]
%    \centering
%    \includegraphics[width=0.8\linewidth]{Figures/HAR_LifecycleMatrix.png}
%    \caption{Representative FM-for-HAR works positioned along the nine taxonomy axes. A checkmark (\checkmark) indicates a substantive contribution to that phase. A blue cross ($X$) marks phases that are out of scope or not required for the work’s stated objective (rather than a deficiency), or only lightly treated. The figure underscores that FM contributions may legitimately target a subset of the constructed phases (e.g., large-scale pretraining, adaptation interfaces, or downstream specialization) without covering the entire lifecycle.}
%    \label{fig:lifecycle_matrix}
%\end{figure}

\begin{table*}[t]
\centering
\caption{Representative foundation models categorized by modality scope.}
\label{tab:modality_scope}
\renewcommand{\arraystretch}{1.12}
\small
\begin{tabular}{p{2.1cm} p{1.6cm} p{2.4cm} p{7.5cm}}
\toprule
\textbf{Model (Ref.)} & \textbf{Scope} & \textbf{Primary Modality} & \textbf{Key Method / Objective} \\
\midrule
RelCon \cite{xu2024relcon} & Unimodal & IMU & Relative contrasitive objective that improves robustness. \\
oneHAR \cite{Wei_2025} & Unimodal & IMU & Universal IMU embeddings with cross-dataset transfer. \\
SelfHAR-700k \cite{yuan2024self} & Unimodal & Accelerometer & Large-scale accelerometer-only SSL (masked/temporal pretext tasks). \\
Pulse-PPG \cite{saha2025pulse} & Unimodal & PPG & Field-pretraining: Raw noisy PPG leveraged as valuable signal; Relative Contrastive Learning: Motif-based distance function for embeddings. \\
HAR-FM \cite{qiu2025towards} & Multimodal & IMU + Physiology & Vector-quantized multimodal SSL with perplexity-aware objectives. \\
MASTER \cite{zhu2024master} & Multimodal & IMU + (other) & Masked-data modeling for cross-sensor consistency. \\
NORMWEAR \cite{luo2024toward} & Multimodal & Physiology (e.g., ECG/PPG) & Spectro-temporal tokenization with ViT-style encoder for multivariate biosignals. \\
MuJo \cite{fritsch2025mujo} & Multimodal & video, language, pose, simulated IMU & multimodal representation methods employing CLIP-like contrastive learning \\
SensorLM \cite{zhang2025sensorlm} & Cross-modal & Wearables + Language & Masked modeling plus sensor–language contrastive alignment for retrieval/captioning. \\
SensorLLM \cite{li2025sensorllm} & Cross-modal & Motion + LLM & Projection and alignment of sensor embeddings with LLM token space for reasoning. \\
IMU2CLIP \cite{moon2023imu2clip} & Cross-modal & IMU + Text & CLIP-style alignment enabling zero-shot retrieval/captioning. \\
FM-Fi~2.0 \cite{weng2025fm} & Cross-modal & RF + Semantic space & Instance-wise feature association with cross-modal distillation. \\
\cite{xue2024leveraging} & Cross-modal & IoT streams + Semantic space & cross-attention to combine a learnable soft prompt; an auxiliary hard prompt that encodes domain knowledge of the IoT sensing task. \\
\bottomrule
\end{tabular}
\end{table*}

\subsection{Modality Scope}
\label{sec:modality-scope}

A central design choice in foundation models in the sensor-based HAR domain is the range of sensing modalities incorporated during pretraining. Unlike vision or language, sensor-based HAR spans highly diverse sources: from body-worn inertial units to ambient RF signals, physiological biosensors, and contextual modalities. We categorize recent work into three levels of modality scope: unimodal, multimodal, and cross-modal foundation models, with representative works listed in Table \ref{tab:modality_scope}.

\paragraph{Unimodal foundation models.}
Unimodal FMs are pretrained on large corpora of a single sensor type and provide universal representations within that modality, without fusing or aligning with other modalities during pretraining, and showing domain robustness across users/devices/placements via augmentations \cite{Wei_2025, hong2024crosshar, xu2024relcon, yuan2024self}. This scope is especially effective when the modality is abundant, standardized, and widely deployed (e.g., accelerometers in smartphones, RF in smart environments). 
%Representative examples include \textbf{oneHAR} \cite{Wei_2025} for large-scale inertial pretraining with cross-dataset generalization; \textbf{CrossHAR} \cite{hong2024crosshar} uses hierarchical SSL across multiple IMU datasets with heterogeneous device characteristics; \textbf{RelCon} \cite{xu2024relcon} describes an IMU-only (1 billion segments from 87,376 participants) pretraining with a relative contrastive objective that improves robustness to user/device variability; and large-scale accelerometer-only SSL leverages massive accelerometer-only data (700,000 person-days), marking the largest unimodal HAR SSL effort to date \cite{yuan2024self} (hereafter \textbf{SelfHAR-700k}).
These models scale readily because training corpora are abundant and standardized, making them efficient to train and strong within their native modality; however, they provide limited contextual grounding and transfer less effectively to semantic or multimodal tasks.

\paragraph{Multimodal foundation models.}
Multimodal FMs integrate multiple sensor streams jointly during pretraining (e.g., IMU+PPG/EDA, IMU+audio, RF+IMU) to learn a single representation for the combined signal, improving context and robustness to missing channels \cite{qiu2025towards,zhu2024master,luo2024toward}. Architectures use early/late fusion, e.g., shared encoders, cross-attention, or a unified encoder–decoder. 
%\textbf{HAR-FM} \cite{qiu2025towards} pretrains across wearable signals with vector-quantized objectives and perplexity-aware regularization; \textbf{MASTER} \cite{Zhu_2025} jointly models multimodal inputs with masked-data reconstruction, learning cross-sensor consistency. Through the pre-training and fine-tuning on 7 multi-modal HAR datasets, MASTER supports 8 modalities (ACC, Gyro, mmWave, WiFi, Skeleton, Lidar, Infrared, and RGB) and 45 human activities; \textbf{NORMWEAR} \cite{luo2024toward} targets multivariate physiological sensing with spectro-temporal tokenization and ViT-style encoders. It is pretrained on a diverse set of physiological signals (including PPG, ECG, EEG, GSR, IMU) and showing exceptional generalizability across 11 public wearable sensing datasets, spanning 18 applications in mental health, body state inference, vital sign estimation, and disease risk evaluation;
These models gain resilience to missing sensors and capture richer context by combining streams, but they can suffer from modality imbalance (e.g., IMU dominates) and synchronization issues, and typically require more complex training.

% \thomas{Check language} 

\paragraph{Cross-modal foundation models.}
Cross-modal FMs are trained to align or translate between different modalities (e.g., IMU $\leftrightarrow$ text\/video) so that information in one can be retrieved or generated from another, bridging sensor modalities with high-level semantics (language/vision) \cite{yang2025visible, imran2024llasa, xue2024leveraging, zhang2025sensorlm, li2025sensorllm, moon2023imu2clip, ray2025txp, weng2025fm}. 
Architectures often use dual encoders with contrastive losses (CLIP-style) or encoder$\rightarrow$LM stacks, enabling zero-shot recognition, natural-language grounding, and explainability.
%\textbf{Visible Light HAR} \cite{yang2025visible} performed cross-modal alignment task between visible light signals and textual descriptions, and proposed a framework that leverages generative LLMs to decode visible light feature representations into human activity descriptions through sequence-to-sequence modeling; 
%\textbf{LLaSA} \cite{imran2024llasa} aligns sensor embeddings with natural-language prompts and outputs; \cite{xue2024leveraging} align IoT streams (IMU, Wi-Fi, mmWave) to a frozen CLIP text encoder via class-prototype prompts;   
%\textbf{SensorLM} \cite{zhang2025sensorlm} aligns wearable signals with language for retrieval and captioning; \textbf{SensorLLM} \cite{li2025sensorllm} extends large language models with motion sensor embeddings, supporting HAR–language bidirectional reasoning; \textbf{IMU2CLIP} \cite{moon2023imu2clip} aligns IMU data with text embeddings in a CLIP-style framework, demonstrating zero-shot retrieval and captioning; \textbf{TxP} \cite{ray2025txp} explicitly pairs pressure sensing with natural language (PressLang text$\leftrightarrow$pressure corpus) and trains both directions (Text2Pressure and Pressure2Text), advancing pressure-based HAR with broader applications and deeper insights into human movement; and \textbf{FM-Fi~2.0} \cite{weng2025fm} maps RF embeddings into multimodal semantic spaces via contrastive distillation. 
By anchoring sensor representations to language or vision, these models gain semantic grounding, enable zero-/few-shot transfer, and support explainable outputs. However, they often depend on paired sensor–language corpora, are sensitive to alignment noise, and incur higher computational costs.

\paragraph{Takeaways.}
Unimodal FMs remain the most scalable and effective within a single sensing channel; multimodal FMs add contextual robustness but incur synchronization and training overhead; and cross-modal FMs deliver semantic grounding and zero-shot behavior at the cost of paired data and compute. The overall trend in HAR is a gradual shift from unimodal pretraining toward cross-modal integration, reflecting a community-wide emphasis on semantic grounding and explainability.

\subsection{Data Landscape: Collected and Generated Corpora}
\label{sec:data-landscape}

Foundation models ultimately inherit their inductive biases from the data on which they are pretrained; the distributional properties, modality balance, and contextual diversity of these corpora determine the kinds of structures and relationships the model internalizes and later transfers to downstream tasks. To make the taxonomy concrete,
we distinguish (\emph{i})in-the-wild collected corpora that reflect real sensing
conditions (devices, placements, sampling rates, demographics) and (\emph{ii}) generated
datasets created by simulation or generative models to expand coverage, balance classes, and
preserve privacy. The datasets collected for each category are available in the accompanying GitHub repository. 
%This split mirrors design choices elsewhere in the taxonomy: corpus scope and tokenization in the \emph{Prepare} phase, objective/architecture coupling in \emph{Pretrain}, and robustness/evaluation protocols in \emph{Deploy}.

% \paragraph{Practically \thomas{What does that mean (practically)?} collected corpora.}
\paragraph{Corpora of sensor data collected in the wild}
The literature shows a consistent backbone of IMU motion datasets (e.g., PAMAP2, WISDM, RealWorld/RWHAR, Opportunity, UCI-HAR, MotionSense, USC-HAD, HHAR, MHEALTH) used for universal representation learning and cross-dataset transfer \cite{Wei_2025,hong2024crosshar,yuan2024self}; health-oriented physiology sets (e.g., WESAD, Sleep-EDF, MIMIC-III/IV, MIT-BIH Arrhythmia, PTB-XL, TUH EEG) that emphasize robustness to noise and clinical relevance \cite{luo2024toward,thapa2024sleepfm,narayanswamy2024scaling}; ambient smart-home deployments (CASAS family) for contextual ADL \cite{tian2025dailyllm}; and smaller but growing use of RF/WiFi/radar and vision/egocentric video for cross-modal alignment \cite{weng2025fm,moon2023imu2clip}.
%Table~\ref{tab:datasets-families} groups these sources by sensing family, providing an empirical backdrop for modality scope and deployment settings.
%Across Tables~\ref {tab:datasets-metadata_1} and~\ref {tab:datasets-metadata_2}, 
Across the datasets tables from our repository, the dataset landscape spans IMU-only corpora, multimodal wearables, large population studies, biosignals, smart-home event streams, and RF/radar sets. Scales vary dramatically, from a few hours and tens of subjects to billions of records and >100k participants, supporting both small-scale method development and population-level pretraining.

% \begin{table}[t]
% \centering
% \caption{Representative dataset families and typical exemplars.}
% \label{tab:datasets-families}
% \renewcommand{\arraystretch}{1.08}
% \small
% \begin{tabular}{p{6.6cm} p{6.6cm}}
% \toprule
% \textbf{Family} & \textbf{Representative datasets} \\
% \midrule
% Wearable IMU / Motion & PAMAP2; WISDM; RealWorld (RWHAR); Opportunity; UCI-HAR; MotionSense; USC-HAD; HHAR; MHEALTH \\
% Physiological / Clinical (ECG/PPG/EDA/EEG/Sleep) & WESAD; Sleep-EDF; MIMIC-III/IV; MIT-BIH Arrhythmia; PTB-XL; TUH EEG \\
% Ambient Smart-Home & CASAS (Aruba, Milan, Kyoto7, Cairo) \\
% RF / WiFi / Radar & mmWave; WiAR; WiFi CSI (various) \\
% Vision / Egocentric (cross-modal) & Ego4D; EPIC-KITCHENS; Kinetics (for alignment) \\
% \bottomrule
% \end{tabular}
% \end{table}

\paragraph{Generated datasets and augmentation.}
Recent work leverages generative foundation models to alleviate data scarcity, class imbalance, and privacy constraints. A representative line of research uses text-to-motion pipelines in which large language models draft structured activity descriptions that condition motion synthesis models to produce virtual IMU streams, expanding coverage without additional human collection and targeting rare or sensitive scenarios \cite{leng2023benefit}. More broadly, cross-modal generation/translation pipelines increase supervisory diversity. For example, sensor–language grounding and bidirectional sensor$\leftrightarrow$text interfaces that create pseudo-captions or weak labels for pretraining and adaptation \cite{zhang2025sensorlm,chen2024sensor2text}. Paired-modality corpora and distillation to proxy teachers (e.g., RF aligned to vision–language spaces) further mitigate limited annotations while preserving modality-specific inference paths \cite{weng2025fm}. In parallel, privacy-aware learning protocols (e.g., federated variants) complement synthetic data by keeping raw traces local while sharing low-leakage updates \cite{bandyopadhyay2025mharfedllm}. While these strategies reduce dependence on hard-to-collect paired datasets and improve low-label regimes, deployment should include calibration and distributional checks (sensor-noise realism, drift, label fidelity) and evaluation under non-IID protocols to ensure gains translate to real devices and users
\cite{hong2024crosshar}.

\subsection{Tokenization and Representation Strategies}

%\begin{figure*}[t]
%  \centering
%  \includegraphics[width=\textwidth]{Figures/HAR_Tokenization.png}
%  \caption{\highLight{Tokenization and representation for sensor-based HAR. Single-stream tokenization converts raw signals into windows, statistical features, spectrograms, or quantized codes; cross-stream scaffolding then synchronizes modalities via positional/meta encodings and performs token fusion or cross-modal projection.}}
%  \label{fig:fm_har_tokenization}
%\end{figure*}

A key step in scaling foundation models for sensor-based HAR is converting continuous sensor streams into machine-readable units for large-scale model training \cite{spathis2023stephardestpitfallsrepresenting}. We use tokenization broadly to include both (\emph{i}) \emph{per-modality token formation}: turning raw IMU/RF/physiology signals into windows, statistical features, spectrogram patches, or quantized codewords, and (\emph{ii}) \emph{cross-modal scaffolding}: synchronizing streams with positional/metadata encodings and then fusing or projecting tokens into shared spaces. Because HAR sensors are multi-rate and often loosely synchronized, 
noisy, and heterogeneous, effective tokenization must balance temporal fidelity, semantic abstraction, and computational budget. 
We therefore group techniques into six families: four single-stream representations and two cross-stream steps that typically follow them, with representative works summarized in Table \ref{tab:tokeniyation}.

\paragraph{Window-based and patch-level segmentation.}
A common first step is to segment continuous streams into fixed-length windows (often overlapping), which serve as atomic tokens for transformers or masked-reconstruction objectives  \cite{xiong2024novel, wieland2025inertial, xu2024relcon, narain2025speech, thapa2024sleepfm, xue2024leveraging}. 
This design supports positional encoding, batching, and parallel attention over long sequences while avoiding unbounded context lengths.

\paragraph{Feature-based aggregation and statistical embeddings.}
When waveform-level modeling is computationally prohibitive, statistical aggregation converts high-rate signals into compact, semantically meaningful tokens at minute or hour resolution \cite{erturk2025beyond, post2025contextllm, post2025contextllm, li2024sensorllm, narayanswamy2024scaling, khasentino2025personal}. 
Such statistical tokens are common in real-world health data pipelines due to interpretability and reduced sequence length \cite{liu2023biosignal}. 

\paragraph{Spectrogram and frequency-domain embeddings.}
Time–frequency representations provide an alternative tokenization pathway that emphasizes rhythmic structure, periodicity, and spectral dynamics in sensor streams—properties that are especially salient for sleep staging, locomotion, and physiological monitoring \cite{luo2024toward, yuan2024self, Liu_2020, liu2021tera}. It trades fine-grained temporal precision for enhanced robustness to noise and temporal misalignment, making them particularly suitable for dense, quasi-periodic sensing modalities.
However, their reliance on fixed windowing and transform parameters can limit flexibility across activities with irregular or highly transient dynamics.

\paragraph{Discrete and quantized sensor tokens.}
Discretizing continuous sensor streams into symbolic or codebook-based tokens provides a critical bridge between time-series sensing and language-centric foundation models ~\cite{qiu2025towards, ansari2024chronos, pillai2025time2lang}.
By mapping real-valued signals into finite vocabularies, quantization enables masked-token objectives, reuse of NLP architectures, and direct interoperability with large language models, which operate natively over discrete sequences \cite{haresamudram2024towards}.

\paragraph{Multimodal alignment and positional encoding.}
Effective sensor-based HAR foundation models must align heterogeneous modalities in both time and representation space, ensuring that tokens remain comparable. Recent approaches emphasize explicit alignment mechanisms (such as shared embedding spaces, structured temporal fields, or sensor–language bridges) to synchronize and contextualize multimodal inputs before reasoning. Representative systems demonstrate that grounding sensor tokens in aligned temporal and semantic coordinates enables cross-modal interpretation and long-horizon reasoning, particularly when integrating sensor streams with language models \cite{li2024sensorllm,yang2025visible,ouyang2024llmsense}.

\paragraph{Token fusion and cross-modal projection.}
This family is about combining or translating already-aligned tokens by means of early/late fusion, cross-attention, or projection into a shared space, with the goal for producing a joint embedding (for recognition) or a cross-modal mapping (for retrieval/captioning/reasoning) \cite{moon2023imu2clip,weng2025fm,chen2024sensor2text}.
By projecting sensor embeddings into semantically grounded spaces (e.g., vision–language or language-only), such models support zero-shot recognition, cross-modal retrieval, and sensor-conditioned text generation.

\begin{table}[t]
\centering
\caption{Representative families of tokenization and representation strategies in foundation models for sensor-based HAR.}
\label{tab:tokeniyation}
\begin{tabular}{p{3.8cm}p{7.2cm}p{2.4cm}}
\toprule
\textbf{Family} & \textbf{Typical Design Choices} & \textbf{Representative Works (Ref.)} \\
\midrule
Window-based / Patch-level & Overlapping fixed-length windows (e.g., 128 samples), enable batching, masking, temporal attention & \cite{hong2024crosshar, thapa2024sleepfm, wieland2025inertial, xiong2024novel, xu2024relcon}\\
Feature-based Aggregation and statistical embeddings & Daily/minutely summaries (e.g., 26-dim) for ViT or statistical learning & \cite{erturk2025beyond, li2024sensorllm, xue2024leveraging, narayanswamy2024scaling, liu2023biosignal}\\
Spectrogram / Frequency Embedding & STFT/wavelet or motif-based frequency learning; ViT or CNN over spectral patches & \cite{luo2024toward, yuan2024self} \\
Discrete / Quantized Tokens & Product quantization, symbolic codebooks, fixed-token sequences for LLMs & \cite{ansari2024chronos, qiu2025towards, pillai2025time2lang}\\
Multimodal Alignment / Positional Encoding & Special tokens, resampling, modality-aware embeddings for aligned token spaces & \cite{li2024sensorllm, yang2025visible, ouyang2024llmsense}\\
Token Fusion / Cross-modal Projection & Sensor-to-language or vision-space mappings via projection/cross-attention & \cite{chen2024sensor2text, moon2023imu2clip, weng2025fm} \\
\bottomrule
\end{tabular}
\end{table}

\paragraph{Takeaways.}
The six families can be understood as two layers in a single pipeline. The first four operate within a modality to convert raw streams into tokens, and the last two families act across modalities and typically follow per-stream tokenization.
In short, groups 1–4 are parallel choices for single-stream token formation, whereas groups 5–6 are sequential cross-stream scaffolding; they should be selected to match the target tasks (pure recognition vs.\ language-grounded reasoning) and deployment constraints.

\subsection{Base Architectures}

%\begin{figure*}[t]
%  \centering
  % Use the PDF if you export one; otherwise keep the PNG below.
  %\includegraphics[width=\textwidth]{figs/HAR_BaseArchitecture.pdf}
%  \includegraphics[width=0.8\textwidth]{Figures/HAR_BaseArchitecture.png}
%  \caption{\highLight{Four base computation graphs for foundation models in sensor-based HAR.}
  %\textbf{Encoder-only stacks} focus on representation learning using a single sensor encoder with lightweight heads for recognition, retrieval, or forecasting.
  %\textbf{Dual encoders} independently embed sensor and text/vision streams and align them via a shared latent projection (CLIP-style) for retrieval/zero-shot transfer. 
  %\textbf{Encoder--decoder stacks} condition a language/multimodal decoder on encoded sensor tokens (cross-attention) to produce captions, rationales, or structured outputs. 
  %\textbf{Language-model stacks} treat sensing as a token sequence using projection/quantization interfaces for forecasting, analysis, and reasoning.
%  }
%  \label{fig:base_architectures}
%\end{figure*}

% Optional compact summary table (kept minimal to avoid encoder details)
\begin{table*}[t]
\centering
\caption{Four base computation graphs with representative works.}
\label{tab:base-arch-3graph}
\begin{tabular}{p{3.0cm} p{8.6cm} p{2.4cm}}
\toprule
\textbf{Computation graph} & \textbf{Core idea} & \textbf{Representative works (Ref.)} \\
\midrule
Encoder-only stacks & Single sensor encoder; pooled embeddings for linear/probe heads; contrastive or masked pretraining; no paired text/vision encoder and no language decoder &
\cite{erturk2025beyond, xu2024relcon, xu2021limu, Wei_2025, zhang2024unimts} \\
Dual encoders & Separate sensor and text/vision encoders; projection into a shared latent; contrastive alignment for retrieval/zero-shot transfer &
\cite{weng2025fm, xue2024leveraging, weng2024large, moon2023imu2clip, matsuishi2025multimodal, lan2025gem} \\
Encoder--decoder stacks & Sensor encoder conditions a causally masked decoder (often LLM) via cross-attention or equivalent; supports captioning, Q\&A, structured generation &
\cite{zhang2025sensorlm, chen2024sensor2text,narayanswamy2024scaling,miao2024spatial,luo2024toward} \\
Language-model stacks & General-purpose LMs (encoder--decoder or decoder-only) operate over sensor tokens or projected embeddings; compatible with time-series FMs and fixed-length token interfaces &
\cite{pillai2025time2lang, bohi2024large, hong2025llm4har, liu2023large, yuan2025leveraging, imran2024llasa}\\
\bottomrule
\end{tabular}
\end{table*}

Table \ref{tab:base-arch-3graph} summarizes four recurring computation graphs that underpin foundation models for sensor-based HAR and lists the representative works. 
These architectures differ primarily in how sensor tokens are encoded, whether and how they are aligned with external modalities (e.g., text or vision), and the role played by language models in downstream inference. 
Backbone choices such as CNN/TCN blocks, Transformer encoders, state-space models (e.g., S4/Mamba), GNNs, or MoE layers instantiate these graphs rather than defining additional architectural families.

\paragraph{Encoder-only stacks (discriminative encoders).}
Encoder-only stacks learn a single sensor encoder that produces pooled embeddings for recognition, retrieval, or forecasting via linear probes or shallow task heads. 
They are typically pretrained with masked or contrastive objectives and emphasize scalability, robustness, and efficient downstream adaptation rather than semantic grounding.
This design is well suited to large, homogeneous sensing corpora, where abundant unlabeled data enables strong representation learning.
Representative systems demonstrate population-scale pretraining and cross-dataset transfer using universal inertial or behavioral embeddings \cite{xu2021limu,xu2024relcon,Wei_2025,yuan2024self,erturk2025beyond,zhang2024unimts}. 
While encoder-only stacks scale efficiently and perform strongly within a modality, they provide limited support for language grounding or cross-modal reasoning.

\paragraph{Dual encoders (retrieval and alignment).}
Dual-encoder architectures independently embed sensor data and text or vision streams, aligning them in a shared latent space using contrastive objectives.
This CLIP-style design enables zero-/few-shot recognition, cross-modal retrieval, and open-set inference by comparing embeddings via similarity metrics rather than task-specific classifiers.
In sensor-based HAR, dual encoders are widely used for RF, inertial, and multimodal alignment, allowing sensor observations to inherit semantics from vision–language models \cite{weng2024large,weng2025fm,moon2023imu2clip,matsuishi2025multimodal,miao2024goat,lan2025gem}.
These models expose reusable embeddings for retrieval and transfer, but typically rely on curated paired data and do not directly support conditional generation.

\paragraph{Encoder--decoder stacks (conditional generation).}
Encoder–decoder stacks pair a sensor encoder with a causally masked decoder (often a language model) that cross-attends to encoded sensor tokens.
This structure explicitly models $\mathbb{P}(Y \mid X)$ and supports conditional generation tasks such as captioning, explanation, Q\&A, or structured output.
Such architectures unify representation learning with grounded generation and are therefore preferred when interpretability, instruction-following, or semantic outputs are required.
In HAR, encoder–decoder stacks are commonly instantiated through masked autoencoding pipelines or sensor-to-language conditioning \cite{chen2024sensor2text,miao2024spatial,narayanswamy2024scaling,luo2024toward}.
The added decoder capacity improves expressiveness but increases computational cost during both pretraining and inference.

\paragraph{Language-model stacks (encoder--decoder or decoder-only).}
Language-model stacks treat sensor observations as token sequences that can be directly consumed by a general-purpose LM after projection or quantization into its embedding space.
Unlike encoder–decoder stacks, the LM itself serves as the primary computation backbone, supporting forecasting, analysis, reasoning, and interactive querying.
This design leverages mature LLM tooling (prompting, instruction tuning, PEFT) and enables rapid task reuse across domains.
Recent HAR systems demonstrate on-device adaptation, health reasoning, and long-horizon behavioral analysis using decoder-only or hybrid LM stacks bridged to sensor tokens \cite{hong2025llm4har,liu2023large,khasentino2025personal,post2025contextllm,yuan2025leveraging}.
While highly flexible, LM stacks depend critically on tokenization quality and may incur higher latency or energy costs.

\paragraph{Takeaways.} 
The four computation graphs trade off scalability, expressivity, and downstream flexibility. Encoder-only stacks excel at bidirectional representation learning and efficient discriminative adaptation (linear heads and adapters); they are the natural fit for masked reconstruction and contrastive pretraining with strong cross-dataset transfer. Dual encoders excel when the goal is alignment and retrieval: they scale to large corpora, enable zero-/few-shot transfer, and keep inference simple by exposing pooled embeddings. Encoder–decoder stacks are preferable when the output must be generative or explanatory (e.g., captioning, rationales, or structured reports), since a decoder can condition on sensor tokens through cross-attention while preserving a strong encoder for discrimination. Language-model stacks (encoder–decoder or decoder-only) treat sensing as a token stream and leverage mature LM tooling for instruction following, reasoning, and interactive outputs, at the cost of careful tokenization and potential latency. In practice, systems increasingly mix these patterns via lightweight projection heads, allowing modular swaps of tokenizers and backbones while retaining a consistent interface to evaluation and downstream adapters.

\subsection{Pretraining Paradigms}
\label{sec:pretraining}

%\begin{figure*}[t]
%  \centering
  % Prefer the vector PDF if you have it; otherwise keep the PNG include.
  %\includegraphics[width=\textwidth]{HAR_Pretraining.pdf}
%  \includegraphics[width=\textwidth]{Figures/HAR_Pretraining.png}
%  \caption{\highLight{Pretraining paradigms for sensor-based HAR.}
  %\textbf{Contrastive} learns cross-view/cross-modal alignment in a shared latent space (CLIP-like), enabling zero-/few-shot transfer and retrieval. 
  %\textbf{Generative} uses masked reconstruction or causal prediction to model temporal continuity and support imputation and text-conditioned decoding. 
  %\textbf{Hybrid / Self-supervised} combines contrastive and generative objectives (often at scale) and introduces semantic grounding via language/distillation, improving robustness across users and devices.
%  }
%  \label{fig:pretraining-paradigms}
%\end{figure*}

\begin{table}[t]
\centering
\caption{Representative pretraining paradigms in foundation models for Human Activity Recognition (HAR). 
Each paradigm emphasizes different pretext objectives, architectural biases, and generalization properties.}
\label{tab:pretraining_paradigms}
\renewcommand{\arraystretch}{1.15}
\small
\begin{tabular}{p{1.6cm} p{3.5cm} p{3.1cm} p{5.8cm}}
\toprule
\textbf{Paradigm} & \textbf{Representative Models} & \textbf{Core Objective} & \textbf{Key Contributions / Characteristics} \\
\midrule

\textbf{Contrastive} &
\cite{xue2024leveraging}, FM-Fi~\cite{weng2024large},  IMU2CLIP~\cite{moon2023imu2clip},   RelCon~\cite{xu2024relcon}&
Cross-view or cross-modal alignment; maximize mutual information between augmented or multimodal sensor pairs &
Distills knowledge from vision/text foundation models into RF or inertial encoders; supports zero-/few-shot HAR via shared embedding space; enhances robustness to device and subject variability. \\

\textbf{Generative} &
LLaSA~\cite{imran2024llasa}, STMAE~\cite{miao2024spatial}, SelfPAB~\cite{logacjov2024selfpab}, Dual-View FM~\cite{wieland2025inertial}, LSM~\cite{narayanswamy2025scaling} &
Masked signal modeling, sequence reconstruction, or causal prediction over time &
Learns temporal continuity and contextual dynamics from unlabeled wearable data; supports interpolation and cross-sensor imputation; enables text-conditioned decoding for semantic reconstruction. \\
\textbf{Hybrid / Self-supervised} &
HAR-DoReMi~\cite{ban2025har},  NORMWEAR~\cite{luo2024normwear},  SensorLLM~\cite{li2025sensorllm}, CrossHAR~\cite{hong2024crosshar}, SensorLM~\cite{zhang2025sensorlm} &
Combines masked prediction, temporal order recovery, and cross-modal contrastive or distillation tasks &
Unifies discriminative and generative learning; scales to millions of hours of wearable data; introduces semantic grounding with natural-language supervision; improves cross-user and cross-device generalization. \\

\bottomrule
\end{tabular}
\end{table}

Pretraining defines how large volumes of unlabeled sensor data are converted into general-purpose representations and is therefore the most critical stage in building foundation models for sensor-based HAR. 
Across the literature, three dominant paradigms have emerged: \emph{contrastive}, \emph{generative}, and \emph{hybrid/self-supervised}, each encoding different inductive biases about temporal structure, modality alignment, and robustness. Table~\ref{tab:pretraining_paradigms} summarizes these paradigms and representative instantiations.

\paragraph{Contrastive pretraining.}
Contrastive approaches learn discriminative embeddings by maximizing agreement between related views (augmentations, devices, or modalities) while separating unrelated samples.
Inspired by \cite{chen2020simple} and CLIP-style \cite{radford2021learning} alignment, contrastive pretraining has become a dominant strategy for cross-domain and cross-modal transfer in HAR.
In sensor-based settings, contrastive objectives align inertial, RF, or multimodal streams either across sensing views or against semantic teachers from vision or language models.
This paradigm emphasizes invariance to user, device, and environmental variation and naturally supports zero-/few-shot recognition and retrieval \cite{weng2024large,moon2023imu2clip}.
Variants such as relative contrastive learning further relax strict positive–negative assumptions, improving robustness under heterogeneous sampling and deployment noise \cite{xu2024relcon}.
Overall, contrastive pretraining provides strong cross-domain generalization but often requires large batches and modality-specific augmentations.

\paragraph{Generative pretraining.}
Generative paradigms replace pairwise discrimination with reconstruction or prediction, using masked modeling or sequence forecasting to learn temporal continuity and contextual structure.
By forcing the model to recover missing segments or predict future windows, these methods capture both local dynamics and long-range dependencies without explicit negative sampling.
Masked autoencoding has proven especially effective for dense, high-frequency modalities such as accelerometers and physiological signals, where temporal coherence is strong \cite{miao2024spatial,logacjov2024selfpab}.
Encoder--decoder formulations extend this idea to population-scale wearable data by modeling long-horizon summaries (e.g., minute- or day-level aggregates) \cite{narayanswamy2024scaling}.
Generative objectives are well suited for denoising, interpolation, and imputation, but often require additional adaptation or alignment to support semantic reasoning or cross-modal transfer.

\paragraph{Hybrid and large-scale self-supervised pretraining.}
Hybrid paradigms combine multiple self-supervised signals, typically masked reconstruction with contrastive alignment, consistency regularization, or robustness-oriented objectives, to balance discriminability, temporal modeling, and scalability.
This convergence reflects a broader trend toward unifying structure learning with semantic grounding.
In sensor-based HAR, hybrid strategies have been shown to improve cross-dataset generalization and robustness under non-IID shifts by integrating hierarchical temporal consistency, physically informed augmentations, or distributionally robust optimization \cite{hong2024crosshar,ban2025har}.
At larger scale, hybrid models incorporate language grounding or instruction-style supervision to align sensor representations with interpretable semantic spaces, enabling bidirectional sensor–language reasoning \cite{zhang2025sensorlm,li2025sensorllm,khasentino2025personal}.
Physiology-focused systems similarly adopt multi-task self-supervision over spectro--temporal tokens to improve robustness across subjects and devices \cite{luo2024toward}.
Overall, hybrid pretraining represents the dominant trajectory of current HAR foundation models, trading architectural simplicity for improved generalization and interpretability.

\paragraph{Takeaways.}

Pretraining paradigms in sensor-based HAR differ not only in objective design but also in the properties they emphasize. Contrastive methods excel at cross-domain alignment and retrieval, generative methods at temporal coherence and noise robustness, and hybrid approaches at scaling these benefits while incorporating semantic grounding. The field is increasingly converging toward hybrid, large-scale self-supervised pretraining as a foundation for adaptable, general-purpose HAR systems, mirroring trends previously observed in vision and language foundation models.

\subsection{Adaptation Strategies}

%\begin{figure*}[t]
%  \centering
%  \includegraphics[width=\textwidth]{Figures/HAR_Adaptation.png}
%  \caption{\highLight{Mechanism-centric view of adaptation in HAR foundation models: PEFT, Full/Partial fine-tuning, and Instruction-tuning \& alignment}}
%  \label{fig:har_adaptation_mechanisms}
%\end{figure*}

% Table for § Adaptation Strategies (mechanism-centric; deployment locus removed)
\begin{table*}[t]
\centering
\caption{Representative adaptation strategies focusing on mechanism-level families and typical design choices with illustrative references.}
\label{tab:adaptation-strategies}
\renewcommand{\arraystretch}{1.12}
\small
\begin{tabular}{p{2.6cm} p{8.6cm} p{2.8cm}}
\toprule
\textbf{Adaptation family} & \textbf{Typical design choices} & \textbf{Representative works (Ref.)} \\
\midrule
Adapters / LoRA / PEFT &
Freeze most or all of the backbone; insert small trainable modules (adapters, LoRA/QLoRA, prefix/soft prompts) on tokenized sensor streams or projection bridges; few-shot updates; low memory/compute for rapid domain adaptation. &
\cite{xiong2024novel, pillai2025time2lang, xie2025physllm, yan2024language, li2025sensorllm} \\
\addlinespace[3pt]
Full or Partial Fine-tuning &
End-to-end or layer-wise unfreezing; selectively update deeper blocks while freezing early features; align projector+head after sensor$\to$LM bridging; specialize compact decoder-only stacks when resources permit. &
\cite{narayanswamy2024scaling, miao2024spatial, hong2025llm4har, xu2021limu, xu2024relcon} \\
\addlinespace[3pt]
Instruction-tuning \& alignment &
Prompt-only in-context alignment (ICL; zero update) \emph{and} supervised formatting (SFT/PEFT) on curated sensor–text exemplars and task templates; aligns prompts, label taxonomies, and output schemas &
\cite{tian2025dailyllm, chen2024sensor2text, zhang2025sensorlm, li2025sensorllm, demirel2025using} \\
\bottomrule
\end{tabular}
\end{table*}

We frame adaptation in sensor-based HAR foundation models by the \emph{mechanism used to modify model behavior}, rather than by task or deployment setting. As summarized in Table~\ref{tab:adaptation-strategies}, existing approaches consistently fall into three mechanism-centric families: (i) Parameter-Efficient Fine-Tuning (PEFT), which keeps the backbone largely frozen while learning small, trainable add-ons; (ii) Full or Partial Fine-Tuning, which updates all or selected layers to increase task-specific plasticity; and (iii) Instruction-Tuning, In-Context Alignment, and Supervised Formatting, which aligns models to task templates and label taxonomies either via prompt-only in-context learning (no weight updates) or via supervised alignment (lightweight finetuning on curated sensor–text exemplars).

\paragraph{Parameter-Efficient Fine-Tuning (PEFT).}
PEFT strategies keep most or all of the pretrained backbone frozen while introducing small trainable components—such as adapters, LoRA/QLoRA modules, or soft prompts—at selected layers or at sensor$\rightarrow$LM projection interfaces.
This design minimizes memory and computation, enabling rapid, label-efficient transfer and making PEFT particularly attractive for personalization, on-device learning, and federated settings.
Across sensor-based HAR, PEFT is commonly applied to bridge time-series encoders and language models or to specialize large backbones to new cohorts with few samples.
Empirically, PEFT balances stability and adaptability: population-level priors are preserved in the frozen backbone, while low-rank updates absorb user-, device-, or task-specific shifts \cite{xiong2024novel,pillai2025time2lang,li2025sensorllm}.
Federated variants further demonstrate that training only adapters or last-layer heads supports privacy-preserving personalization without sharing raw sensor data \cite{bandyopadhyay2025mharfedllm}.
Overall, PEFT has emerged as the default adaptation mechanism when compute, labels, or data access are constrained.

\paragraph{Full or partial fine-tuning.}
When domain shift is substantial or tighter coupling between sensing features and task objectives is required, full or selective fine-tuning remains effective.
Rather than updating all parameters indiscriminately, many HAR systems adopt \emph{partial} fine-tuning—freezing early layers while unfreezing deeper blocks or projection heads—to balance plasticity and generalization.
This strategy is frequently used when adapting pretrained encoders to new sensor placements, populations, or task semantics under moderate supervision \cite{xu2021limu,yuan2024self}.
At scale, reconstruction-pretrained backbones can be fine-tuned on annotated activity or health tasks to specialize representations for deployment \cite{narayanswamy2024scaling}.
Compact decoder-only stacks further enable selective fine-tuning for efficient on-device specialization \cite{hong2025llm4har}.
While more costly than PEFT, full or partial fine-tuning remains a strong baseline when sufficient labels and resources are available.

\paragraph{Instruction-tuning and alignment.}
Instruction-tuning governs \emph{how tasks are specified and outputs are structured}, rather than which parameters are updated.
In sensor-based HAR, instruction alignment spans two complementary regimes:
(\emph{i}) \emph{prompt-only in-context learning} (ICL), where task templates, label taxonomies, and few-shot exemplars steer inference without weight updates; and
(\emph{ii}) \emph{supervised alignment} via SFT or PEFT on curated sensor–text pairs to improve format adherence, consistency, and faithfulness.
Prompt-only alignment enables rapid zero-shot reuse across tasks and datasets, especially when sensor streams are summarized into structured textual descriptions \cite{thapa2025stressllm,cleland2024leveraging}.
Supervised instruction-tuning is typically applied when stable output schemas, multi-turn interaction, or grounded explanation is required, as in sensor–language stacks that couple encoders to language decoders \cite{chen2024sensor2text,zhang2025sensorlm}.
Crucially, instruction-tuning is orthogonal to PEFT or full fine-tuning: it specifies the supervision interface and can be combined with either mechanism to improve usability without extensive retraining.

\paragraph{Takeaways.}
The three families delineate how HAR foundation models are adapted. PEFT is typically preferred when labels or compute are scarce (or on-device updates are needed): the backbone remains frozen while small modules supply just enough capacity to absorb domain shift. Full/partial fine-tuning becomes attractive when distribution mismatch or new sensing characteristics demand deeper representational change and sufficient supervision is available. Instruction-tuning \& alignment is complementary and spans two regimes: (i) prompt-only in-context learning (zero parameter updates) for fast task formatting, and (ii) supervised formatting via SFT/PEFT on curated sensor–text exemplars to stabilize outputs and enforce label taxonomies. These mechanisms can be composed (e.g., instruction-tuned PEFT). For example, the authors in \cite{xu2025exploring} evaluate zero-/few-shot in-context learning (ICL) with chain-of-thought, and then apply parameter-efficient fine-tuning (LoRA) as instruction-tuning on curated instruction–answer pairs, substantially improving few-shot accuracy.

\subsection{Downstream Capabilities}
\label{sec:downstream-capabilities}

% === Figure: Downstream Capabilities & Generalization Protocols ===
%\begin{figure*}[b]
%  \centering
  % Replace the filename below with your actual asset name (pdf/png).
  % Vector PDF is preferred for camera-ready.
%  \includegraphics[width=\textwidth]{Figures/HAR_DownstreamCapabilities.png}%
%  \caption{\highLight{Downstream capabilities and the accompanying generalization protocols used in sensor-based HAR. }
%Top row (left→right): \textit{Zero-/few-shot \& open-set}: recognize unseen activities with $k$-shot label budgets and open-set rejection; 
%\textit{Cross-dataset / device / user}: train on dataset~A and test on unseen datasets/devices/users with leave-one-out and cross-position splits; 
%\textit{Cross-modal retrieval \& search}: sensor↔text/video retrieval evaluated by Recall@K and mAP under cross-domain splits with a shared embedding space. 
%Bottom row: \textit{Captioning, Q\&A, reasoning}: sensor-conditioned decoding (prompts/PEFT), measured by caption/Q\&A accuracy and human/expert ratings; 
%\textit{Reconstruction, forecasting, imputation}: masked-reconstruction/denoising and short/long-horizon forecasting under distribution shift; 
%\textit{Federated \& on-device evaluation}: client-level personalization with communication rounds, reporting edge latency/energy and privacy constraints.
%}
%  \label{fig:downstream_capabilities}
%\end{figure*}

\begin{table*}[t]
\centering
\caption{Downstream capability families and their typical \emph{generalization protocols}. Citations are chosen to be disjoint across rows.}
\label{tab:downstream-capabilities}
\renewcommand{\arraystretch}{1.12}
\small
\begin{tabular}{p{3.3cm} p{7.2cm} p{3.6cm}}
\toprule
\textbf{Capability family} & \textbf{Protocol focus} & \textbf{Representative works (Ref.)} \\
\midrule
Zero-/few-shot \& open-set recognition & Unseen activities; open-set thresholds; few-shot label budgets & \cite{weng2024large, weng2025fm, li2025zara, civitarese2025large, ji2024hargpt} \\
Cross-dataset/device/user generalization & Train on dataset A, test on unseen dataset/device/user; leave-one-subject-out & \cite{qiu2025towards, Wei_2025, zhu2024master, xiong2024novel, yuan2024self} \\
Cross-modal retrieval \& search & Sensor$\leftrightarrow$text/video retrieval (recall@K, mAP); cross-domain splits & \cite{matsuishi2025multimodal, choube2025gloss, li2025vital, fang2024physiollm, thapa2024sleepfm} \\
Language-grounded captioning / Q\&A / reasoning & Caption correctness, Q\&A accuracy, human studies; prompt robustness & \cite{zhang2025sensorlm, chen2024sensor2text, demirel2025using, fukazawa2025llm, ugwu2025potential} \\
Generative reconstruction / forecasting / imputation & Masked reconstruction error; denoising; future-window forecasting & \cite{miao2024spatial, hong2024crosshar, liurobusthar, zhang2024unimts, narayanswamy2024scaling} \\
Federated \& on-device evaluation & Client-level personalization; communication rounds; edge latency/energy & \cite{bandyopadhyay2025mharfedllm, zhang2024enabling, hong2025llm4har, bukit2025activity, tian2025dailyllm} \\
\bottomrule
\end{tabular}
\end{table*}

Once pretrained or adapted, HAR foundation models expose a spectrum of downstream capabilities that are typically evaluated under specific generalization protocols. We focus on \emph{what the model can do} and \emph{how it is evaluated}. Table~\ref{tab:downstream-capabilities} summarizes this landscape by organizing downstream use into six recurring capability families:

\paragraph{Zero-/few-shot \& open-set recognition.}
This capability evaluates label efficiency and open-world robustness by testing whether pretrained representations can recognize unseen or sparsely labeled activities without retraining. Most approaches rely on alignment priors (typically cross-modal or semantic embedding spaces) to support zero-/few-shot inference and open-set rejection. Distillation from vision–language teachers enables strong open-set RF recognition~\cite{weng2024large, weng2025fm}, while knowledge-augmented or retrieval-based reasoning supports zero-shot motion understanding~\cite{li2025zara}. More broadly, recent work probes the extent to which general-purpose LLMs can serve as zero-shot recognizers when paired with structured prompts, in-context learning, or tool-augmented reasoning~\cite{civitarese2025large, ji2024hargpt, kim2024eeg}.

\paragraph{Cross-dataset / cross-device / cross-user generalization.}
This capability measures robustness under non-IID shifts by training on one (or multiple) datasets and evaluating on unseen datasets, devices, or users. Foundation models improve generalization by learning sensor-invariant representations through large-scale, heterogeneous pretraining, masked or vector-quantized objectives, and lightweight adaptation interfaces. Empirical evidence shows that multimodal pretraining, universal IMU embeddings, and population-scale self-supervision consistently reduce performance collapse under cross-domain transfer~\cite{qiu2025towards, Wei_2025, zhu2024master, xiong2024novel, yuan2024self}.

\paragraph{Cross-modal retrieval and search.}
This capability evaluates whether sensor representations can be aligned with text or vision in a shared embedding space, enabling sensor$\leftrightarrow$text/video retrieval under cross-domain splits. Foundation models typically realize this via contrastive or distillation-based alignment, allowing activity traces to be queried and indexed across modalities without task-specific retraining~\cite{matsuishi2025multimodal, fang2024physiollm}. Performance is commonly measured using retrieval metrics such as Recall@K or mAP, reflecting semantic grounding quality rather than classification accuracy~\cite{thapa2024sleepfm}. These shared embeddings support practical workflows including search over passive sensing logs~\cite{choube2025gloss} and expert-in-the-loop longitudinal analysis across heterogeneous sensors and cohorts~\cite{li2025vital}.

\paragraph{Language-grounded captioning, Q\&A, and reasoning.}
This capability evaluates whether sensor representations can be translated into faithful natural-language descriptions, answers, or rationales, thereby exposing semantic structure beyond label prediction. Most systems pair a sensor encoder with a language decoder or LM bridge to generate captions and responses grounded in wearable or ambient signals~\cite{chen2024sensor2text, zhang2025sensorlm}. Two-stage alignment pipelines map sensor embeddings into a language model’s representation space, enabling both sensor-conditioned generation and text-guided recognition over a shared backbone~\cite{li2025sensorllm}. Token-level interfaces that compress time series into fixed-length embeddings further allow frozen LLMs to reason over sensor traces without full retraining~\cite{pillai2025time2lang}. Cross-modal decoding is also extended to non-traditional sensing, such as visible-light dynamics mapped to activity descriptions~\cite{yang2025visible}. Long-horizon reasoning is supported by encoding explicit temporal fields and metadata alongside sensor tokens, improving prompt stability and explanatory coherence~\cite{ouyang2024llmsense}. 
%Evaluation typically emphasizes semantic grounding—via caption accuracy, Q\&A correctness, or human judgments—highlighting interpretability and generalization rather than classification accuracy alone.

\paragraph{Generative reconstruction, forecasting, and imputation.}
Generative capabilities focus on modeling temporal continuity and recoverability via masked reconstruction, denoising, or future-window forecasting. Evaluation typically reports reconstruction error or forecasting skill under held-out spans and distribution shifts, rather than label accuracy. Representative approaches include spatial–temporal masked autoencoding \cite{miao2024spatial}, hierarchical masked modeling for transfer \cite{hong2024crosshar}, population-scale masked autoencoding on aggregated wearables \cite{narayanswamy2024scaling}, causal and masked modeling across placements~\cite{wieland2025inertial}, and multi-scale robustness-oriented SSL~\cite{liurobusthar}, ~\cite{zhang2024unimts}. These protocols emphasize temporal coherence and robustness under limited supervision.

\paragraph{On-device, federated, and online adaptation.}
Deployment-centric evaluations assess whether foundation models can adapt after deployment while respecting privacy and resource constraints. This includes on-device fine-tuning, federated personalization, and online or test-time adaptation. Compact decoder-only stacks demonstrate efficient on-device specialization ~\cite{hong2025llm4har}, while daily-life assistants coordinate local inference with lightweight adaptation ~\cite{tian2025dailyllm}. Federated approaches with LLM components enable user-level personalization without centralizing raw data ~\cite{bandyopadhyay2025mharfedllm}, and context-aware agent frameworks orchestrate privacy-aware reasoning across devices ~\cite{post2025contextllm}. Smartphone personalization~\cite{zhang2024enabling} and semi-supervised federated learning for activity transitions ~\cite{bukit2025activity} further illustrate this setting. Metrics emphasize personalization gain, communication cost, on-device latency/energy, and privacy leakage alongside accuracy.

\paragraph{Takeaways.} Downstream capabilities reflect how pretraining choices cash out in practice: alignment-heavy FMs are suitable for zero-/few-shot recognition and cross-modal retrieval; generative objectives deliver strong reconstruction, imputation, and forecasting; and language-bridged stacks enable captioning, Q\&A, and reasoning. Robustness is ultimately determined by the evaluation protocol, not just accuracy on an IID split, but also performance under cross-dataset/device/user shifts, as well as open-set conditions. For deployability, federated and on-device protocols surface privacy, latency, and energy trade-offs that offline metrics can hide. A sound reporting recipe is to pair each claimed capability with the matching generalization protocol and metric (e.g., recall@K for retrieval, open-set F1 for zero-shot, reconstruction error/forecast skill for generative tasks). Practically, choose the lightest capability path that meets the application’s constraints, reserving heavier cross-modal reasoning and online adaptation for settings that truly need semantic grounding or continual robustness.

\subsection{Deployment Settings}
\label{sec:deployment-settings}

%\begin{table*}[t]
%\centering
%\caption{Deployment settings emphasized by recent sensor-based HAR foundation models}
%\label{tab:deployment_settings}
%\renewcommand{\arraystretch}{1.12}
%\small
%\begin{tabular}{p{2.0cm} p{5.6cm} p{4.0cm} p{2.0cm}}
%\toprule
%\textbf{Setting} & \textbf{Typical design choices} & \textbf{Protocols \& deployment metrics} & \textbf{Representative works} \\
%\midrule
%\textbf{Cloud-scale training \& centralized evaluation} &
%Large-scale pretraining/alignment on centralized servers; contrastive distillation to vision/language teachers; masked/autoencoding over population-scale corpora; heavy backbones with full fine-tuning; inference in centralized environment. &
%Cross-dataset/device/user splits; open-set/zero-/few-shot tests; retrieval metrics (Recall@K, mAP); reconstruction/forecast skill; standardized centralized test sets. &
%\cite{weng2024large, weng2025fm, narayanswamy2024scaling} \\
%\midrule
%\textbf{On-device \& mobile execution} &
%Edge-friendly tokenizers; parameter-efficient fine-tuning (adapters/LoRA/QLoRA); quantization and pruning; projection bridges to compact LMs; selective unfreezing; privacy-by-design; streaming-friendly inference. & On-device latency, energy, and memory; real-time throughput; local personalization effectiveness; occasional cloud sync (if any). &
%\cite{xue2024leveraging, ouyang2024llmsense, yan2024language, zhang2024enabling, hong2025llm4har, leng2024imugpt} \\
%\bottomrule
%\end{tabular}
%\end{table*}

\begin{table*}[t]
\centering
\caption{Deployment settings emphasized by recent sensor-based HAR foundation models}
\label{tab:deployment_settings}
\renewcommand{\arraystretch}{1.12}
\small
\begin{tabular}{p{2.0cm} p{8.6cm} p{2.4cm}}
\toprule
\textbf{Setting} & \textbf{Typical design choices} & \textbf{Representative works} \\
\midrule
\textbf{Cloud-scale training \& centralized evaluation} &
Large-scale pretraining/alignment on centralized servers; contrastive distillation to vision/language teachers; masked/autoencoding over population-scale corpora; heavy backbones with full fine-tuning; inference in centralized environment. &
\cite{weng2024large, weng2025fm, narayanswamy2024scaling} \\
\midrule
\textbf{On-device \& mobile execution} &
Edge-friendly tokenizers; parameter-efficient fine-tuning (adapters/LoRA/QLoRA); quantization and pruning; projection bridges to compact LMs; selective unfreezing; privacy-by-design; streaming-friendly inference. &
\cite{xue2024leveraging, ouyang2024llmsense, yan2024language, zhang2024enabling, hong2025llm4har, leng2024imugpt} \\
\bottomrule
\end{tabular}
\end{table*}

Beyond architecture and pretraining, \emph{where a foundation model runs} fundamentally shapes its evaluation and practical relevance in sensor-based HAR. Across the literature, two deployment regimes dominate (Table~\ref{tab:deployment_settings}): cloud-scale training with centralized evaluation, and on-device or mobile execution under resource constraints.

\paragraph{Cloud-scale training and centralized evaluation.}
Cloud-centric HAR foundation models pretrain or align large backbones on centralized infrastructure to exploit massive, heterogeneous corpora and establish strong representational priors under controlled evaluation protocols. Typical designs emphasize cross-modal distillation (e.g., aligning RF or wearable signals to vision–language teachers) and large-scale masked or autoencoding objectives over population-level aggregates to maximize cross-dataset robustness and open-set transfer~\cite{weng2024large, weng2025fm, narayanswamy2024scaling}.

\paragraph{On-device and mobile execution.}
On-device sensor-based HAR foundation models redesign the path from sensor tokens to predictions to satisfy tight latency, energy, and privacy constraints. Common strategies include structured temporal encoding for stable LLM reasoning over long sensor traces, edge–cloud cooperation with local open-set filtering, and decoupling semantic reasoning from lightweight inference backbones~\cite{ouyang2024llmsense, xue2024leveraging, yan2024language}. In parallel, practical deployments rely on compact adaptation techniques (quantization, PEFT, pruning, and selective unfreezing) to localize personalization while bounding memory and compute~\cite{zhang2024enabling, hong2025llm4har}. Lightweight encoder-only designs further demonstrate that few-shot recognition can be achieved by mapping sensor activations to text-compatible representations without full language decoding~\cite{cruciani2025few}. 
%Collectively, these systems converge on a shared design pattern: minimal tokenization overhead, small task-specific adaptation heads, and frozen or partially frozen backbones that preserve accuracy under mobile execution budgets.

\paragraph{Takeaways.}
Cloud-scale and on-device settings provide complementary evaluation lenses rather than competing solutions. Centralized pipelines reveal what sensor-based HAR foundation models \emph{can learn}, whereas edge-centric studies reveal what they \emph{can sustain} in practice. Meaningful comparison therefore requires pairing non-IID robustness metrics with system-level costs on target hardware. In practice, future HAR foundation models will likely combine cloud-scale pretraining with lightweight, privacy-aware on-device adaptation.

\subsection{Application Domains}
\label{sec:application-domains}

% === Figure: Application Domains (icons-only wheel) ===
\begin{figure*}[t]
  \centering
  % Place the exported file in your figs/ directory; vector PDF preferred.
  \includegraphics[width=0.6\textwidth]{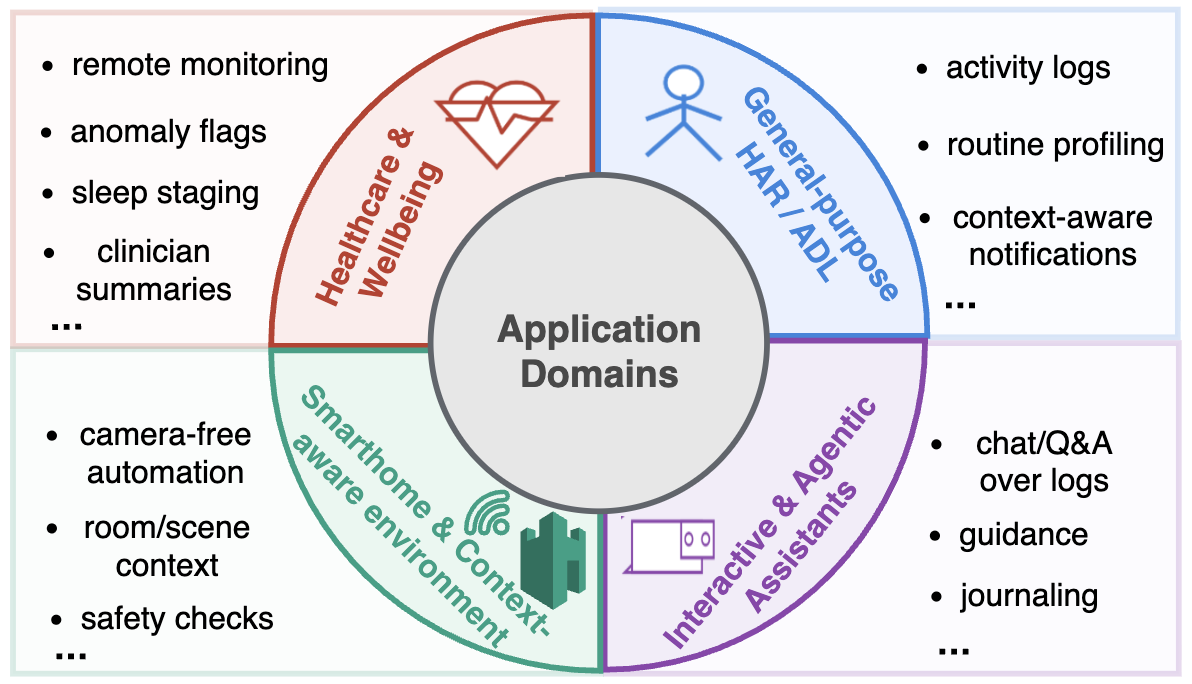}
  \caption{Application domains for sensor-based HAR foundation models. 
  The radial layout highlights four commonly targeted areas: general-purpose HAR / daily living \cite{xue2024leveraging, Wei_2025,hong2024crosshar,qiu2025towards} , 
  healthcare and wellbeing \cite{xie2025physllm,li2025vital,saha2025pulse,thapa2024sleepfm,bohi2024large}, smart-home and context-aware environments \cite{weng2024large,yang2025visible,yang2025contextagent, mahmoudi2024leveraging} , and 
  interactive/agentic assistants \cite{ouyang2024llmsense,yan2024language, tian2025dailyllm}.}
  \label{fig:application-domains-wheel}
\end{figure*}

%\begin{table*}[t]
%\centering
%\caption{Representative \emph{application domains} in HAR domain foundation models.}
%\label{tab:application_domains}
%\renewcommand{\arraystretch}{1.12}
%\small
%\begin{tabular}{p{3.6cm} p{6.6cm} p{3.2cm}}
%\toprule
%\textbf{Domain} & \textbf{Typical sensors / tasks} & \textbf{Representative works (Ref.)} \\
%\midrule
%\textbf{General-purpose HAR / ADL} &
%Smartphone/waist-worn IMU; multi-dataset daily-activity recognition; cross-dataset/device/user transfer; robustness to placement heterogeneity &
%\cite{xue2024leveraging, Wei_2025,hong2024crosshar,qiu2025towards} \\
%\midrule
%\textbf{Healthcare \& wellbeing (incl.\ sleep)} &
%Physiology streams (ECG/PPG/EDA, temperature) and wearables for clinical/wellbeing use; sleep staging and nocturnal routines; retrieval/summarization and language-grounded interpretation &
%\cite{xie2025physllm,li2025vital,saha2025pulse,thapa2024sleepfm,bohi2024large} \\
%\midrule
%\textbf{Smart home \& context-aware environments} &
%Ambient RF or visible-light sensing; device-free activity recognition; cross-room/scene generalization; camera-free privacy-aware setups &
%\cite{weng2024large,yang2025visible,yang2025contextagent} \\
%\midrule
%\textbf{Interactive \& agentic assistants} &
%Multi-sensor streams + LLMs; annotation, captioning, Q\&A and daily assistance; on-device prompting and personalization &
%\cite{ouyang2024llmsense,yan2024language, tian2025dailyllm} \\
%\bottomrule
%\end{tabular}
%\end{table*}

Foundation models in the sensor-based HAR domain are being deployed across a spectrum of real-world settings. Based on the surveyed works, four areas stand out (as depicted in  Fig. \ref{fig:application-domains-wheel}): (i) general-purpose HAR and activities of daily living (ADL), (ii) healthcare \& wellbeing (including sleep), (iii) smart-home and context-aware environments, and (iv) interactive/agentic assistants:

\paragraph{General-purpose HAR / ADL.}
General-purpose sensor-based HAR targets everyday activities in natural settings and underpins common mobile and wearable applications, including time-use summaries, context-aware notifications, fitness and wellness tracking, and routine profiling across devices and placements. Foundation models in this domain aim to provide a single, reusable backbone that remains robust to real-world variability such as missing channels, device heterogeneity, and user-dependent motion patterns. Large-scale inertial and multimodal pretraining has shown that such backbones can generalize across datasets and sensing configurations without task-specific redesign~\cite{Wei_2025, hong2024crosshar}. Vector-quantized and multimodal self-supervised objectives further improve resilience under cross-device and cross-context evaluation, supporting broad ADL coverage in consumer and health-facing deployments~\cite{qiu2025towards}.

\paragraph{Healthcare \& wellbeing.}
Healthcare-oriented sensor-based HAR leverages physiological sensing and language grounding to support continuous monitoring, clinical interpretation, and personalized feedback. Foundation models in this domain emphasize robustness to noise, motion artifacts, and inter-subject variability while enabling higher-level reasoning over long-term biosignal trends. Language-augmented pipelines integrate physiological priors with LLM-based inference for clinically aligned interpretation and explanation~\cite{xie2025physllm}, while multimodal retrieval and summarization frameworks support longitudinal health analytics across heterogeneous sensor streams~\cite{li2025vital}. Signal-centric pretraining further improves resilience under real-world acquisition conditions, such as motion- and illumination-corrupted photoplethysmography~\cite{saha2025pulse}. Sleep analysis emerges as a canonical application, where masked signal modeling enables label-efficient staging from wearable data~\cite{thapa2024sleepfm}, and signal-to-text mappings facilitate clinically interpretable summaries of nocturnal physiological patterns~\cite{bohi2024large}.

\paragraph{Smart-home \& context-aware environments.}
In smart-home and context-aware environments (those with event-based sensor streams), foundation models operate on fundamentally different data characteristics compared to wearable sensing. Typical deployments, such as CASAS-style environments, rely on binary, event-driven sensor streams (e.g., motion detectors[on/off], door sensors[open/close], etc.), where each signal encodes sparse state transitions rather than continuous measurements \cite{zhao2026deep}. These signals are asynchronous, low-frequency, and highly sparse, often representing when an interaction occurs without directly capturing motion dynamics. This leads to unique challenges, including temporal sparsity, ambiguity of sensor activations (i.e., the same sensor pattern may correspond to different activities), and strong dependence on contextual and routine information. These properties require substantially different modeling and pre-training strategies for foundation models. Rather than learning fine-grained motion representations, FMs must capture long-term dependencies, event co-occurrence structures, and semantic routines across time. This motivates representations based on discrete event tokens, long-horizon sequence modeling, and integration of contextual or language-based priors to resolve ambiguity. Emerging work has begun to explore this direction by leveraging large language models (LLMs) for zero-shot and early-stage activity recognition in such smart-home environments. For example, recent studies demonstrate that LLMs can map sparse sensor event sequences into natural language descriptions and achieve competitive zero-shot performance without labeled data, particularly during the cold-start phase of deployment \cite{hiremath2024game}. Moreover, hybrid approaches that combine LLMs with supervised models enable a transition from zero-shot inference to data-driven adaptation as labeled data accumulate, highlighting a promising paradigm for continual activity recognition in sparse sensing environments \cite{10607364}. However, compared to foundation models in wearable HAR, these approaches remain limited in scale and generality. Current methods largely rely on prompt engineering, rule-based sensor-to-language conversion, or hybrid pipelines, rather than unified large-scale pretraining on ambient event streams. As such, the development of foundation models specifically tailored to sparse, event-driven smart-home data remains at an early stage. Beyond recognition, recent efforts further extend toward context-aware reasoning by aligning ambient sensing modalities (e.g., RF or visible light) with semantic representations~\cite{weng2024large,yang2025visible}. These directions suggest a shift from activity classification toward long-horizon contextual intelligence, where foundation models act not only as recognizers but also as reasoning engines over ambient environments.

\paragraph{Interactive \& agentic assistants.}
Interactive foundation-model pipelines operationalize sensor representations as user-facing actions, enabling natural-language queries, reflective feedback, and context-aware guidance over personal activity streams. Long-horizon reasoning over structured sensor traces supports multi-turn explanation and self-reflection (e.g., querying behavioral changes around inactivity) without requiring task-specific retraining~\cite{ouyang2024llmsense}. Lightweight, on-device language grounding further enables motion snippets to be rendered as concise textual cues for reminders and prompts, avoiding cloud dependency while preserving privacy~\cite{yan2024language}. At the system level, daily-life assistants emphasize routine summarization, anomaly flagging, and proactive nudging through on-device prompting and aggregation, shifting HAR from passive recognition toward continuous, agentic support~\cite{tian2025dailyllm}.

\section{Major Directions in Foundation Models in HAR domain}
\label{sec:major-directions}

The taxonomy in the last section describes \emph{what} types of models exist and how they differ technically and operationally; however, it does not yet describe the developmental trajectories and research philosophies that guide current progress.  This section, therefore, organizes the landscape into three overarching directions that collectively define the evolution of HAR foundation models: 
%(1)~developing HAR-specific foundation models from scratch on large-scale sensor corpora, (2)~adapting general time-series or multimodal foundation models to the HAR domain, and (3)~leveraging large language models (LLMs) for sensor understanding and reasoning. These directions are complementary to the taxonomy: each draws on different subsets of the technical and operational axes introduced in Section~\ref{sec:taxonomy}, together forming a unified picture of how the community is progressing from data-driven sensor modeling toward semantically grounded, human-centric foundation models for ubiquitous computing.

\subsection{Developing HAR-Specific Foundation Models from Scratch}
\label{sec:har-specific-fm}

The first major direction in the evolution of foundation models in the sensor-based HAR domain is the development of domain-native models trained entirely from raw sensor streams. Rather than adapting architectures from language or vision, this line of research builds foundation models from scratch using large corpora of wearable, ambient, and physiological data, aiming to learn generalizable, sensor-aware representations. Some representative works are listed in Table \ref{tab:har_specific_fm}.

\begin{table*}[t]
\centering
\caption{Representative sensor-native foundation models developed from scratch for Human Activity Recognition.}
\label{tab:har_specific_fm}
\begin{tabular}{p{2.1cm} p{2.3cm} p{2.2cm} p{2.2cm} p{4.5cm}}
\toprule
\textbf{Model (Ref.)} & \textbf{Learning Paradigm} & \textbf{Modality} & \textbf{Scale / Dataset} & \textbf{Key Contributions} \\
\midrule
HAR-FM \cite{qiu2025towards} & Hybrid SSL (Contrastive + Quantization) & IMU + Physiology & Cross-user wearable datasets & Quantization-based SSL with perplexity diversity regularizer for heterogeneous sensors \\
MASTER \cite{zhu2024master} & Masked Reconstruction & Multimodal (IMU + PPG + EDA) & Multi-sensor corpora & Multimodal masked modeling capturing cross-sensor dependencies \\
NORMWEAR \cite{luo2024normwear} & Masked SSL + Fusion & Physiology (PPG + EDA + ECG) & Health-oriented datasets & Physiological FM enabling anomaly and stress prediction \\
oneHAR \cite{Wei_2025} & Contrastive + Cross-Dataset SSL & IMU & Multi-dataset HAR corpora & Universal IMU FM for dataset-agnostic representation learning \\
LSM \cite{narayanswamy2025scaling} & Empirical Scaling Study & Multimodal Wearables & Large public datasets & Establishes scaling laws and performance plateaus for wearable FMs \\
\bottomrule
\end{tabular}
\end{table*}

\paragraph{Motivation and significance.}

Training from scratch on heterogeneous sensor streams allows the model to directly capture long-range temporal dependencies, user variability, and biomechanical correlations without domain translation.  
Recent studies report substantial gains in cross-dataset transfer when pretraining directly on sensor corpora (e.g., IMU-focused and physiology-focused FMs outperform general time-series baselines across multiple benchmarks \cite{qiu2025towards,Wei_2025,luo2024normwear}).  
Moreover, such models serve as universal feature extractors for personalization, health monitoring, and multimodal reasoning tasks.

\paragraph{Learning paradigms and scaling behaviors.}
Most HAR-specific FMs employ a combination of masked reconstruction and contrastive alignment to balance temporal continuity and discriminative separability.  
Generative objectives (e.g., masked signal modeling, forecasting) improve robustness to missing data, whereas contrastive or distillation objectives ensure inter-user and inter-device invariance.  
Hybrid formulations \cite{qiu2025towards, zhu2024master}, yield particularly strong cross-domain transfer capability.  
Inspired by scaling analyses in physiological FMs \cite{pillai2024papagei}, LSM \cite{narayanswamy2025scaling} confirms that models trained on smaller datasets exhibit limited generalization capacity, whereas scaling up to 10\^{}8 parameters results in significant gains in test loss and generative zero-shot performance. These findings highlight the need to align model size with adequate data to fully leverage the model’s representational power

\paragraph{Data diversity and domain heterogeneity.}
Building robust HAR FMs requires exposure to diverse motion contexts, body locations, and sensor configurations. While vision and language FMs benefit from billions of web-scale samples, the HAR community relies on distributed, small-scale datasets with non-IID characteristics. Approaches such as synthetic data generation through physics-based simulation \cite{oishi2025physically, ray2025improving, xu2023practically} and generative models \cite{rey2019let, leng2025agentsense, leng2023generating, alamgeer2025ai} are increasingly explored to fill this gap. Moreover, multi-device harmonization pipelines (e.g., signal normalization, sampling alignment, and domain-adversarial training) play a key role in enabling cross-platform generalization.

\paragraph{Challenges and opportunities.}
Developing sensor-native foundation models remains constrained by limited public datasets, privacy regulations, and computational costs of long-sequence pretraining. Addressing these issues demands, first, an open, privacy-preserving sensor data consortia; second, efficient architectures such as sparse transformers and memory-compressed encoders; and third, multimodal integration with contextual modalities (language, audio, video) for semantic grounding.  Despite these challenges, the steady emergence of large-scale models \cite{zhu2024master, qiu2025towards} demonstrates that domain-specific pretraining is no longer optional but foundational for scalable, interpretable, and human-centric activity modeling.

\subsection{Adapting General Time-Series and Multimodal Foundation Models to HAR}
\label{sec:adapt-har}

\begin{table*}[t]
\centering
\caption{Representative approaches that adapt general time–series or multimodal foundation models to HAR. 
%We list the mechanism used to adapt the pretrained backbone, the base FM being reused, sensor/semantic modalities, and the primary contribution.
}
\label{tab:adapt_har}
\renewcommand{\arraystretch}{1.12}
\small
\begin{tabular}{p{2.0cm} p{2.4cm} p{2.2cm} p{1.8cm} p{4.8cm}}
\toprule
\textbf{Model (Ref.)} & \textbf{Adaptation Strategy} & \textbf{Base Foundation Model} & \textbf{Modality} & \textbf{Key Contribution} \\
\midrule
%Chronos HAR Adapters \cite{xiong2024novel} & Adapters / PEFT on frozen backbone & Chronos (general time-series FM) & IMU & Lightweight adapters trained with forecasting-style SSL, enabling efficient few-shot transfer and on-device use while preserving Chronos knowledge. \\
%\addlinespace[3pt]
IMU2CLIP \cite{moon2023imu2clip} & Projection + contrastive alignment & CLIP (vision–language FM) & IMU $\leftrightarrow$ text/video & Maps IMU segments into CLIP space for zero-shot activity recognition and cross-modal retrieval with frozen CLIP weights. \\
\addlinespace[3pt]
Time2Lang \cite{pillai2025time2lang} & Sensor\,$\to$\,LM token bridge & Chronos (TS FM) + LLaMA (LLM) & Wearables + language & Projects variable-length time series to fixed-length token matrices so a largely frozen LLM can reason over sensor traces. \\
\addlinespace[3pt]
%FM-Fi \cite{weng2024large} & Cross-modal distillation & ViT/CLIP teacher (vision/VL FM) & RF (+ video/text for training) & Distills semantics from vision/VL teachers into an RF encoder; RF-only inference with strong cross-device generalization. \\
%\addlinespace[3pt]
FM-Fi~2.0 \cite{weng2025fm} & Contrastive distillation + projection & CLIP-like multimodal teacher & RF $\leftrightarrow$ semantic space & Aligns RF embeddings to multimodal semantic space with shared projection; improves retrieval and zero-/few-shot transfer. \\
\addlinespace[3pt]
AURA\textendash MFM \cite{matsuishi2025multimodal} & Multibranch alignment with reused backbones & Pretrained video/text encoders (e.g., CLIP-like) & IMU + video + mocap + text & Couples an IMU encoder with pretrained vision/text branches; shared space enables multimodal retrieval and recognition. \\
\addlinespace[3pt]
GEM \cite{lan2025gem} & ECG branch + multimodal alignment & Pretrained vision/language components & ECG + image + text & Unifies ECG time series with images and language via alignment heads, leveraging existing VL components for clinician-aligned interpretation. \\
%\addlinespace[3pt]
%IoT Sensing \cite{xue2024leveraging} & Prompt learning + partial fine-tuning with supervised contrastive alignment & Frozen CLIP-style text encoder (general multimodal FM) & IMU, WiFi/mmWave (IoT RF) & Dual-encoder sensor↔text alignment with soft+hard class-prototype prompts for zero-/open-set HAR; optional GAN-based augmentation for unseen classes \\
%Speech FMs $\rightarrow$ Wearables \cite{narain2025speech} & Probing + LoRA (PEFT) & HuBERT / wav2vec\,2.0 (speech FMs) & IMU, ECG, PPG & Cross-domain reuse of speech encoders; upsampling/pooling windowing; early-layer transfer; probes/LoRA yield strong HAR and health predictions. \\
\bottomrule
\end{tabular}
\end{table*}

A second major direction in the development of foundation models for sensor-based HAR leverages the rapid progress of general-purpose time-series and multimodal foundation models. Instead of training sensor-native models entirely from scratch, this approach adapts pre-existing foundation backbones to the characteristics of human-motion and wearable data.  This direction emphasizes efficiency, transferability, and data scalability: by reusing pretrained representations from large generic corpora, researchers can overcome the scarcity of labeled HAR data while achieving strong generalization to new users and devices. Some representative works are listed in Table \ref{tab:adapt_har}.

\paragraph{Motivation and rationale.}
General time-series foundation models such as Chronos \cite{ansari2024chronos}, TimeGPT \cite{garza2023timegpt}, and PatchTST \cite{nie2022time} have shown that large-scale pretraining on diverse sequential data (financial, weather, IoT) yields representations transferable to other temporal domains.  
For sensor-based HAR, where sensor data are also structured time-series with repetitive patterns, these pretrained models provide a powerful initialization that accelerates convergence and improves low-data performance. Adaptation further reduces energy cost and training time compared to full-scale pretraining, enabling deployment on embedded or edge devices.

\paragraph{Performance and generalization.}
Empirical studies indicate that adapting general time-series or multimodal foundation backbones to HAR can achieve competitive performance at substantially lower training cost, especially under low-label regimes \cite{xiong2024novel,moon2023imu2clip,pillai2025time2lang,chen2024sensor2text,weng2025fm}. When combined with domain-specific adapters or small-scale fine-tuning, such models show strong cross-dataset and cross-subject transfer \cite{xiong2024novel}. 
%For instance, \textbf{Chronos HAR Adapters}~\cite{xiong2024novel} replace full fine-tuning with small adapter layers on a frozen Chronos backbone, substantially reducing on-device update overhead while maintaining accuracy comparable to full retraining.

\paragraph{Challenges and outlook.}
Although adaptation provides efficiency and scalability, it faces limitations in semantic grounding and modality alignment. Pretrained models rarely contain motion-specific inductive biases, and their time-series priors may not capture high-frequency sensor noise or multi-rate sampling typical in wearables.  Future work should focus on multimodal adaptation combining generic time-series FMs with sensor-language or sensor-vision encoders, and on bridging representation gaps through cross-modal distillation.  
Ultimately, adaptation represents a pragmatic step toward universal “foundation backbones” for time-series data, where sensor-based HAR acts both as a benchmark and a beneficiary.

\subsection{Leveraging Large Language Models for Human Activity Recognition}
\label{sec:llm-har}

\begin{table*}[t]
\centering
\caption{Representative LLM–based approaches for sensor-based Human Activity Recognition.}
\label{tab:llm_har}
\begin{tabular}{p{1.8cm} p{1.7cm} p{2.3cm} p{3.0cm} p{4.8cm}}
\toprule
\textbf{Model (Ref.)} & \textbf{Role of LLM} & \textbf{Modality} & \textbf{Learning Paradigm} & \textbf{Key Contributions} \\
\midrule
%PH\mbox{-}LLM~\cite{khasentino2025personal} & Coach/reasoner & Wearables+ Language(sleep, HR/HRV, steps/activity) & Instruction-tuning (SFT) + multimodal adapter alignment; hybrid supervised & Health coaching with expert-level reasoning; rubric-based evaluation of explanations 
%\\
%SensorLM \cite{zhang2025sensorlm} & Semantic encoder & Wearables + Language & Masked modeling + sensor–language contrastive & Zero-shot recognition, retrieval, and cross-modal grounding \\
%SensorLLM \cite{li2025sensorllm} & Bidirectional reasoning engine & Motion Sensors + LLM & Fine-tuned multimodal transformer & Interactive natural-language prompting and explanation \\
DailyLLM \cite{tian2025dailyllm} & Generative explainer & Multimodal (IMU + PPG + context) & Autoregressive LLM integration & Generates daily activity logs and personalized summaries \\
HARGPT \cite{ji2024hargpt} & Zero-shot recognizer & IMU (tokenized) & Prompt-based LLM reasoning & Treats sensor data as language for open-world recognition \\
%MHARFedLLM \cite{bandyopadhyay2025mharfedllm} & Federated personalized model & Multimodal Wearables & Federated multimodal training & LLM-based personalization for cross-user adaptation \\
PhysLLM \cite{xie2025physllm} & Health interpreter & Physiology + Language & Cross-modal embedding alignment & Extends LLM reasoning to remote physiological sensing \\
LanHAR \cite{yan2024language} & Language explainer & IMU + Language & Sensor-to-text translation & Generates natural-language activity descriptions for smart environments \\
ContextLLM \cite{post2025contextllm} &
Reasoning engine &
Wearables + ambient (multi-device aggregation) &
Prompt-only (zero-shot; no fine-tuning) &
Hierarchical phone-hub aggregation; robustness analysis under misclassification;\\
%(Smart Home) \cite{cleland2024leveraging} & Zero-/few-shot labeler; supervised text classifier & Ambient binary sensors & Prompting + BERT SFT over sensor-to-text prompts& Sensor$\to$text pipeline; prompt engineering and compression; analysis of label confusions/hallucinations. \\
%LLaSA~\cite{imran2024llasa} & captioning / Q\&A  & IMU + Text & Self-supervised sensor encoder + projector; LoRA instruction-tuning & SensorCaps and OpenSQA datasets; zero-shot HAR on seen/unseen sets; language-grounded reasoning over IMU. \\
StressLLM~\cite{thapa2025stressllm} & Prompted predictor  & Wearables/ physiology  & In-context learning over windowed/aggregated features & biomarker ablations and age/gender sensitivity analysis \\
%EEG\mbox{-}GPT \cite{kim2024eeg} & classification and tool-augmented reasoning & EEG & SFT on verbalized features; zero-shot with ICL & zero-shot above chance with ICL; transparent, stepwise reasoning via specialist EEG tools \\
\bottomrule
\end{tabular}
\end{table*}

The third major direction in foundation model research for sensor-based HAR explores how LLMs can enhance, interpret, or even perform activity recognition tasks.  Unlike Sections~\ref{sec:har-specific-fm} and~\ref{sec:adapt-har}, which emphasize sensor-specific pretraining or general time-series and multimodal FMs adaptation, this direction investigates LLMs as multimodal reasoning engines that connect sensor signals with natural language, enabling semantic understanding, explainability, and human–AI interaction. Recent works demonstrate that LLMs can serve as powerful bridges between sensor representations and human-level reasoning by translating raw time-series into structured descriptions, questions, or decisions. A few representative works are listed in Table \ref{tab:llm_har}.

\paragraph{Motivation and conceptual shift.}
The integration of LLMs into HAR stems from the growing realization that physical sensor signals encode human behavior that is inherently semantic. Traditional classifiers output discrete activity labels, but fail to represent complex routines, context, or intent. LLMs, pretrained on massive textual corpora, possess compositional reasoning and contextual understanding that can translate sensor-derived embeddings into interpretable activity descriptions. By incorporating language priors, HAR moves from label prediction toward narrative, context-aware activity understanding.

\paragraph{Language-grounded multimodal reasoning.}
LLM-driven HAR models share a common goal: aligning low-level sensor embeddings with high-level semantic concepts. Contrastive alignment links motion patterns with language embeddings \cite{sensorlm2025}, while generative or autoregressive alignment treats sensor streams as language tokens \cite{tian2025dailyllm, ji2024hargpt}. Recent hybrid methods further embed LLMs within multimodal transformer pipelines, combining visual, linguistic, and sensor modalities for holistic contextual reasoning \cite{yan2024language,luo2024normwear}. This multimodal convergence echoes biosignal-oriented architectures such as PhysLLM \cite{xie2025physllm}, which integrate physiological features with textual representations for remote health monitoring.

\paragraph{Challenges and open questions.}
Despite rapid progress, sensor–language integration poses substantial challenges.  
First, large-scale paired sensor–text corpora remain scarce; existing datasets (e.g., PAMAP2, WISDM) lack descriptive annotations.  
Second, aligning continuous sensor data with discrete linguistic tokens introduces modality gaps and synchronization noise.  
Third, the computational and privacy overhead of deploying LLMs in edge or federated environments limits real-world applicability.  
Emerging directions aim to overcome these issues via synthetic caption generation, hierarchical sensor summarization, and teacher–student distillation between FMs and LLMs.  
Moreover, as HAR expands toward multimodal wellbeing modeling, LLM-driven agents could eventually reason about both activities and their implications, enabling proactive, human-centered ubiquitous systems.

%\paragraph{Outlook.}
%The integration of LLMs into HAR marks a shift from perception to cognition: from detecting human actions to interpreting human behavior.  
%By coupling sensor intelligence with natural language understanding, LLM-based FMs promise generalization, transparency, and accessibility, paving the way for next-generation HAR systems that can converse, explain, and adapt in natural human terms.

%\section*{Summary and transition.}
\subsection*{Summary}
The three major directions outlined in this section collectively chart the current evolution of foundation models in the sensor-based HAR domain. Developing sensor-native models from scratch establishes the technical and conceptual foundation for learning generalizable motion representations directly from large-scale sensor corpora. Adapting general time-series or multimodal FMs offers a pragmatic pathway toward efficiency and scalability, enabling knowledge transfer from broader temporal domains. Finally, the integration of large language models extends HAR from perception to cognition, embedding semantics, reasoning, and explainability into activity understanding. Together, these directions are complementary rather than competitive: domain-native pretraining provides the representational base, cross-domain adaptation expands reach and efficiency, and LLM integration unlocks semantic grounding and interactive intelligence. As the field matures, future research will likely converge on hybrid frameworks that unify these three directions by combining large-scale sensor pretraining, modular adaptation, and language-grounded reasoning within a single foundation model.

\section{Open Challenges and Future Directions}
\label{sec:future-directions}

Despite rapid progress, foundation models (FMs) for sensor-based HAR remain at an earlier stage of maturity than their vision and language counterparts. 
The central challenges are no longer confined to improving accuracy on isolated datasets, but instead concern how to scale, adapt, evaluate, and responsibly deploy general-purpose sensor representations.
Below we distill the most pressing research directions that are specific to the foundation-model paradigm in sensor-based HAR.

\subsection{Scaling Sensor Pretraining under Fragmentation and Privacy Constraints}

A defining bottleneck for HAR foundation models is the absence of large, unified corpora comparable to ImageNet or web-scale text.
Unlike vision or language, sensor data are inherently fragmented across devices, sampling rates, placements, and populations, and are tightly coupled to privacy concerns.
Recent works show that pretraining on heterogeneous sources improves robustness (e.g., FM-Fi~2.0~\cite{weng2025fm}, HAR-FM~\cite{qiu2025towards}, LSM~\cite{narayanswamy2024scaling}), but they also expose the limits of current aggregation strategies.

This raises FM-specific questions that go beyond classical dataset collection: how to learn transferable priors from non-IID, partially observed sensor streams; how to balance population-scale pretraining with user-level variability; and how to exploit synthetic or simulated data without degrading downstream behavior.

\subsection{Representation Limits: Temporal Hierarchy and Semantic Abstraction}

While foundation models promise reusable representations, HAR exposes fundamental gaps in how current models encode long-horizon temporal structure and semantic abstraction.
Human activities unfold across nested time scales (seconds to days) and are only weakly anchored to linguistic concepts, making direct transfers from vision--language paradigms nontrivial.

Recent models begin to address parts of this challenge through masked temporal modeling~\cite{miao2024spatial, narayanswamy2024scaling} or language-aligned representations~\cite{zhang2025sensorlm, li2025sensorllm}, but a principled treatment of hierarchical temporal representations remains underdeveloped.
Key open questions include how to encode routines and transitions in a form that remains stable across users and devices, and how to align sensor embeddings with language without overfitting to prompts or captions.
These challenges point to new pretraining objectives that explicitly model temporal hierarchy and abstraction.

\subsection{Adaptation and Deployment as First-Class FM Concerns}

In HAR, where and how a foundation model adapts is as important as what it learns during pretraining.
Unlike cloud-only vision FMs, HAR systems generally should operate under tight constraints on latency, energy, personalization, and privacy.
Recent work on parameter-efficient adaptation and edge deployment~\cite{xiong2024novel, hong2025llm4har, ouyang2024llmsense} demonstrates feasibility, but also highlights the fragility of naive fine-tuning under distribution shift.

This elevates adaptation and deployment to first-class FM research problems.
Open directions include designing adaptation interfaces that support few-shot personalization without catastrophic drift, coupling cloud-scale pretraining with on-device or federated updates, and co-designing models with sensing hardware and edge accelerators.
Addressing these issues requires treating adaptation not as a downstream engineering step, but as a core component of the FM lifecycle.

\subsection{Evaluation, Responsibility, and Interactive Behavior}

Finally, foundation models demand a rethinking of evaluation and responsibility. Accuracy on IID test splits is insufficient for models intended to generalize, adapt, and interact with users in the wild. Recent studies reveal both the promise and risk of language-mediated sensing, particularly in health-related contexts~\cite{kim2024eeg, bandyopadhyay2025mharfedllm}. Beyond recognizing robustness and reliability as key challenges, it is essential to clarify how these properties are evaluated in the context of foundation models for HAR. Unlike conventional HAR systems that primarily report accuracy on IID benchmarks, FM evaluation emphasizes performance under distribution shifts and open-world conditions. In practice, robustness is assessed through cross-domain protocols, including cross-user, cross-device, and cross-dataset generalization, as well as resilience to noise, missing data, and modality variations. Reliability further requires models to produce trustworthy and interpretable outputs. This can be evaluated through confidence calibration, uncertainty estimation, and the ability to detect out-of-distribution or novel activities. For language-integrated or agentic HAR systems, additional aspects such as robustness to hallucination and consistency of reasoning become critical.

These considerations highlight several open challenges: establishing standardized protocols for cross-dataset and open-set evaluation, reporting system-level costs (latency, energy, memory) alongside accuracy, and ensuring robustness against hallucination and privacy leakage in language-grounded outputs. As HAR FMs increasingly support interactive and agentic behavior~\cite{ji2024hargpt, tian2025dailyllm}, responsible deployment will require calibrated uncertainty, verifiable reasoning processes, and human-in-the-loop oversight. Therefore, a comprehensive evaluation of HAR foundation models should jointly consider (i) generalization performance under non-IID conditions, (ii) robustness to perturbations and incomplete sensing, and (iii) reliability of predictions and model confidence—moving beyond conventional accuracy-centric benchmarks.

\paragraph{Outlook.}
Taken together, these challenges suggest that the next phase of sensor-based HAR foundation models will be defined less by scaling backbones and more by advances in pretraining objectives, lifecycle-aware adaptation, deployment-conscious design, and rigorous evaluation.
Addressing these issues is essential for transforming foundation models from promising research artifacts into reliable, human-centered sensing systems.

\section{Conclusion}
\label{sec:conclusion}

Foundation Models are redefining the paradigm of sensor-based HAR, shifting the focus from narrow, dataset-specific classifiers toward general-purpose, semantically grounded, and human-centric representations. This survey has provided the first comprehensive overview of this emerging field, covering both the technical and operational taxonomy of current approaches and the conceptual directions shaping future research. Through an integrated analysis of recent literature, we have discussed how sensor-native pretraining, cross-domain adaptation, and language-grounded reasoning collectively enable the transition from perception to cognition in HAR. 
We further highlighted the open challenges that must be addressed to ensure responsible and efficient deployment. Looking ahead, we foresee a convergence of the three major trajectories described in this paper: (1) scalable sensor-native pretraining for robust generalization, (2) modular adaptation for efficient personalization and deployment, and (3) multimodal integration with LLMs for interpretability and contextual intelligence. Together, these directions point toward the emergence of a new generation of Sensor Foundation Models which is capable of reasoning, explaining, and interacting with the physical world through ubiquitous sensing. Such models will not only advance scientific understanding of human behavior but also pave the way for responsible, human-aligned artificial intelligence in everyday life.

\begin{acks}
The authors used the large language model ChatGPT-4 to proofread the text and improve its grammatical flow. All generated feedback was critically assessed and validated by the authors. The final content and conclusions are the sole responsibility of the authors.
\end{acks}

\paragraph{Ethical, Legal, and Social Considerations.}
Foundation models for sensor-based HAR increasingly operate in regulated and socially sensitive contexts, particularly when applied to health monitoring, behavioral analysis, or continuous personal sensing. 
Emerging regulatory frameworks, such as the EU AI Act, highlight these systems as potentially high-risk and emphasize requirements for transparency, data governance, and human oversight. 
While this survey does not aim to provide a legal or regulatory analysis, we acknowledge that such frameworks reinforce the importance of privacy-by-design, responsible data handling, and accountable deployment of HAR foundation models. These considerations motivate ongoing research into interpretable models, privacy-preserving learning, and deployment-aware evaluation practices.

\bibliographystyle{ACM-Reference-Format}
\bibliography{sample-base}

\end{document}